\documentclass[twocolumn]{aastex63}

\usepackage{graphicx}	
\usepackage[T1]{fontenc}
\usepackage{multirow}
\usepackage{amsmath}
\newcommand{\msun}{\mathrm{M_\odot}}
\shorttitle{dark and stellar mass around isolated central galaxies}
\shortauthors{Alonso et al.}
\graphicspath{{./}{figures/}}

\begin{document}

\title{Dark against luminous matter around isolated central galaxies: a comparative study between modern surveys and IllustrisTNG}

\author{Pedro Alonso}
\affiliation{Department of Astronomy, Shanghai Jiao Tong University, Shanghai 200240, China}
\affiliation{Shanghai Key Laboratory for Particle Physics and Cosmology, Shanghai 200240, China}

\author{Wenting Wang}
\affiliation{Department of Astronomy, Shanghai Jiao Tong University, Shanghai 200240, China}
\affiliation{Shanghai Key Laboratory for Particle Physics and Cosmology, Shanghai 200240, China}

\author{Jun Zhang}
\affiliation{Department of Astronomy, Shanghai Jiao Tong University, Shanghai 200240, China}
\affiliation{Shanghai Key Laboratory for Particle Physics and Cosmology, Shanghai 200240, China}

\author{Hekun Li}
\affiliation{Department of Astronomy, Shanghai Jiao Tong University, Shanghai 200240, China}
\affiliation{Shanghai Key Laboratory for Particle Physics and Cosmology, Shanghai 200240, China}

\author{Shi Shao}
\affiliation{Key Laboratory for Computational Astrophysics, National Astronomical Observatories, Chinese Academy of Sciences, Beijing 100101, China}

\author{Qi Guo}
\affiliation{Key Laboratory for Computational Astrophysics, National Astronomical Observatories, Chinese Academy of Sciences, Beijing 100101, China}

\author{Yanqin He}
\affiliation{Tianjin Astrophysics Center, Tianjin Normal University, Tianjin 300387, China}

\author{Cai-Na Hao}
\affiliation{Tianjin Astrophysics Center, Tianjin Normal University, Tianjin 300387, China}

\author{Rui Shi}
\affiliation{Department of Astronomy, Shanghai Jiao Tong University, Shanghai 200240, China}
\affiliation{Shanghai Key Laboratory for Particle Physics and Cosmology, Shanghai 200240, China}

\correspondingauthor{Wenting Wang}
\email{wenting.wang@sjtu.edu.cn}



\begin{abstract}
Based on independent shear measurements using the DECaLS/DR8 imaging data, we measure the weak lensing signals around isolated central galaxies (ICGs) from SDSS/DR7 at $z\sim0.1$. The projected stellar mass density profiles of satellite galaxies are further deduced, using photometric sources from the Hyper Suprime-Cam (HSC) survey (pDR3). The signals of ICGs $+$ their extended stellar halos are taken from \citet{2021ApJ...919...25W}. All measurements are compared with predictions by the IllustrisTNG300-1 simulation. We find, overall, a good agreement between observation and TNG300. In particular, a correction to the stellar mass of massive observed ICGs is applied based on the calibration of \cite{2013ApJ...773...37H}, which brings a much better agreement with TNG300 predicted lensing signals at $\log_{10}M_\ast/\msun>11.1$. In real observation, red ICGs are hosted by more massive dark matter halos, have more satellites and more extended stellar halos than blue ICGs at fixed stellar mass. However, in TNG300 there are more satellites around blue ICGs at fixed stellar mass, and the outer stellar halos of red and blue ICGs are similar. The stellar halos of TNG galaxies are more extended compared with real observed galaxies, especially for blue ICGs with $\log_{10}M_\ast/\msun>10.8$. We find the same trend for TNG100 galaxies and for true halo central galaxies. The tensions between TNG and real galaxies indicate that satellite disruptions are stronger in TNG. In both TNG300 and observation, satellites approximately trace the underlying dark matter distribution beyond $0.1R_{200}$, but the fraction of total stellar mass in TNG300 does not show the same radial distribution as real galaxies. 
\end{abstract}



\section{Introduction} \label{section:Introduction}

Most of the matter in our Universe is in the form of invisible dark matter. Dark matter can collapse to form cosmic sheets, filaments and clumpy halos. The growth of large and intermediate scales of cosmic structures can be successfully modeled by the linear perturbation theory, while on small scales, galaxies form through the gas cooling and condensation within dark matter halos \citep[e.g.][]{1978MNRAS.183..341W}. Smaller halos and galaxies can merge with larger halos, becoming the so-called substructures and satellite galaxies, orbiting around the central dominant galaxy in the host halo. These satellite galaxies would eventually be disrupted by tidal forces, becoming part of the central galaxy. The stripped material from these satellites mainly builds the outskirts of the central galaxy, forming the extended diffuse stellar halos. Understanding the distribution of galaxies on small scales and their connections to the host dark matter halos is one of the key topics in astrophysics, but it is not trivial, since it is beyond the capability of the linear perturbation theory and the baryonic physics complicates the understanding.

A few empirical approaches have been developed to model the formation and evolution of galaxies under the standard framework of cosmic structure formation, including halo occupation distribution (HOD) models \citep[e.g.][]{2000MNRAS.318.1144P, 2000MNRAS.318..203S, 2001ApJ...546...20S, 2002ApJ...575..587B, 2005ApJ...633..791Z}, semi-analytical models (SAMs) \citep[e.g.][]{1991ApJ...379...52W, 1993MNRAS.264..201K, 2001MNRAS.328..726S, 2007MNRAS.375....2D, 2011MNRAS.413..101G} and hydrodynamical simulations such as Illustris \citep[e.g.][]{2014Natur.509..177V, 2014MNRAS.444.1518V, 2014MNRAS.445..175G, 2015MNRAS.452..575S}, EAGLE \citep[e.g.][]{2015MNRAS.446..521S, 2015MNRAS.450.1937C}, and IllustrisTNG \citep[e.g.][]{2017MNRAS.465.3291W, 2018MNRAS.473.4077P, 2018MNRAS.475..648P, 2018MNRAS.480.5113M, 2018MNRAS.477.1206N, 2018MNRAS.475..676S, 2018MNRAS.475..624N}. 

HOD is mathematical, which models the probability of finding an $N$ number of galaxies in a halo of mass $M$, through the probability distribution $P(N|M)$, with the free parameters tuned to match the clustering of observed galaxies. HOD widely assumes that the spatial distribution of satellites trace the dark matter distribution. While more complicated HOD models, such as those considering how the occupation number depends on the colors of the central galaxies, have been developed \citep[e.g.][]{2009MNRAS.392.1080S}, physical mechanisms behind galaxy formation and evolution are not directly considered. SAM solves the non-linearity of galaxy formation by analytical approximations and by tracing the merger histories of galaxies, with the merger trees either generated through Monte-Carlo methods or from dark matter only simulations. Freedoms in adopted physical processes are tuned to match real observations at redshift $z=0$ or higher redshifts. SAMs are computationally cheap, but they have the drawback of depending on assumptions about the physics behind galaxy formation. 

Compared with HOD and SAM, numerical simulations are more sophisticated approaches to study the galaxy-halo connection, although they are more computationally expensive. With the inclusion of baryonic physics, dark matter only simulations have evolved into more complex hydrodynamical simulations. Further developments in the formation and evolution of galaxies include detailed treatments of various physical processes, such as cooling, star formation and evolution, chemical enrichment and gas recycling. The spatial distribution of star particles and gas cells can be well resolved. However, many physical processes adopted in hydrodynamical simulations are still below the resolution limit and have to rely on assumptions of subgrid physics. 

Given the uncertainties of the adopted physical processes, it is essential to perform studies that directly compare and verify that these simulations are as close as possible to the real Universe. Such comparisons are useful in terms of guiding to which direction the models are supposed to be improved, whether the included physics are reasonable and whether there are missing ingredients \citep[e.g.][]{2011MNRAS.417..370G,2012MNRAS.427..428G,2012MNRAS.424.2574W,2014MNRAS.442.1363W,Han15,2017MNRAS.469.2626H,2020MNRAS.495.4570M,2021MNRAS.500..432A,2022MNRAS.517.3579Z}. 

The IllustrisTNG set of simulations are one of the most up-to-date hydrodynamical simulations of galaxy formation and evolution. It has been shown in previous studies that TNG predictions show reasonable agreement with real data, in terms of the global stellar mass and luminosity functions, global color distribution, galaxy clustering and satellite abundance etc.  \citep[e.g.][]{2018MNRAS.475..624N,2018MNRAS.475..676S,2018MNRAS.473.4077P}. However, in terms ,of a few detailed and stringent comparisons, they still deviate from real data, especially when the comparison is made against morphology, the outer stellar halos of massive galaxies and when in combination with weak lensing measurements. For example, \cite{2020MNRAS.495.4570M} compared a few galaxies from the Dragonfly Nearby Galaxies Survey with TNG mass-matched counterparts, and found that real galaxies have less mass or light at large radii, which they denote as a so-called ``missing outskirts problem". More recently and in combination with weak lensing measurements, \cite{2021MNRAS.500..432A} reported that at the virial mass of $\sim 10^{14}\msun$, TNG shows excess in the outer stellar mass. Moreover, \cite{2020MNRAS.498.5804R} claimed that TNG300 tends to predict $\sim$50\% higher lensing signals for red galaxies with $10.2<\log_{10}M_\ast/\msun<11.2$. 

In a previous study, \citet{2021ApJ...919...25W} measured the projected radial density profiles of stellar mass in isolated central galaxies, in their extended stellar halos, and in surviving bound satellite galaxies. The fractions of total stellar mass versus total mass as measured from weak lensing signals are further calculated based on the HSC internal S19 shear measurements, though the weak lensing signals are noisy due to the small footprint. In this study, we first improve the weak lensing measurements using a much larger shear catalog based on galaxy images from the Dark Energy Camera Legacy Survey \citep[DECaLS;][]{2019AJ....157..168D} and the \textsc{Fourier\_Quad} Method. We then measure the projected stellar mass density profiles for satellites following the method of \citet{2021ApJ...919...25W}, with a new threshold in the stellar mass of satellites to ensure fair comparisons with TNG predictions. The projected stellar mass density profiles for centrals and their diffuse stellar halos are directly taken from \citet{2021ApJ...919...25W}. We compare these profiles as well as their connections to each other with the corresponding predictions by IllustrisTNG. 

In particular, the \textsc{Fourier\_Quad} Method \citep[]{2015JCAP...01..024Z,2017ApJ...834....8Z,2019ApJ...875...48Z} is a shear measurement pipeline that calculates the shear in Fourier space. An advantage of this method is that it does not require any assumption about the galaxy morphology. In \cite{2017ApJ...834....8Z}, a novel shear measurement approach was introduced, the so called PDF symmetrization method, which recovers the shear signal by symmetrizing the PDF of the shear estimators. \citet[]{2017ApJ...834....8Z} proved that with this new approach, \textsc{Fourier\_Quad} was allowed to reach the Cramer-Rao Bound, which is the lower bound to the statistical variance of an estimator.

For observational calculations throughout this paper, we adopt as our fiducial cosmological model the first-year Planck cosmology \citep{2014A&A...571A..16P}, with $H_0=67.3\mathrm{km s^{-1}/Mpc}$, $\Omega_\mathrm{m}=0.315$ and $\Omega_\Lambda=0.685$, to be consistent with \citet{2021ApJ...919...25W}. 

The structure of this paper is as follows. We introduce our observational and simulation data in section~\ref{section:data}. In section~\ref{section:Methods}, we describe the \textsc{Fourier\_Quad} Method and introduce our methods of calculating the lensing signals, the projected stellar mass density profiles for satellite galaxies, and extended stellar halos centered on isolated central galaxies. We present our results in section~\ref{section:Results}, and include discussions in section~\ref{sec:disc}. We conclude in section~\ref{section:Conclusion}.

\section{Data} \label{section:data}

In this paper, we will present measurements of the weak lensing signals and the projected stellar mass density profiles for photometric satellite galaxies and for central galaxies $+$ their extended stellar halos. Our sample of central galaxies are selected as those galaxies which are the brightest within a certain volume, which we call isolated central galaxies, and all the measurements are centered on these isolated central galaxies binned according to stellar mass or color. Except for the projected stellar mass profiles for isolated central galaxies and their stellar halos, which were measured by stacking galaxy images in \cite{2021ApJ...919...25W}, and we directly take the measurements from the previous study, all the other measurements are performed independently in this study. All observational measurements will be compared with the predictions by IllustrisTNG300 in detail. In the following, we first introduce the IllustrisTNG suite of simulations, we then move on to describe our selection of isolated central galaxies in both observation and TNG300, followed by descriptions about the weak lensing data and the photometric galaxies from the Hyper Suprime-Cam Imaging Survey used to calculate the projected density profiles for satellite galaxies. 

\subsection{IllustrisTNG}
\label{section:Simulations/IllustrisTNG}

The IllustrisTNG project\footnote{https://www.tng-project.org} \citep[]{2018MNRAS.475..624N} is composed by a series of magneto-hydrodynamical simulations, run with the moving-mesh code AREPO \citep[]{2019ascl.soft09010S}. These simulations are the descendants of the original Illustris simulations, with improved galaxy formation models  \citep{2018MNRAS.480.5113M,2018MNRAS.477.1206N,2018MNRAS.475..624N,2018MNRAS.473.4077P,2018MNRAS.475..676S,2019ComAC...6....2N}, which describe the coupled evolution of gas, dark matter, stars, and black holes. It includes extensive treatments of various galaxy formation and evolution processes, including metal line cooling, star formation and evolution, stellar feedback, chemical enrichment, gas recycling and magneto-hydrodynamics. The simulations were carried out using the Planck 2015 $\Lambda$CDM cosmological model with parameters $\Omega_\mathrm{m}=0.3089$, $\Omega_\Lambda=0.6911$, $\Omega_\mathrm{b}=0.0486$, $\sigma_8=0.8159$, $n_\mathrm{s}=0.9667$, and $h=0.6774$ \citep{Planck2015}. Dark matter halos are identified with the friends-of-friends (FoF) algorithm \citep{Davis1985}. In each FoF group, substructures including galaxies are identified with the SUBFIND algorithm~\citep{SUBFIND}.

IllustrisTNG includes a total of 18 simulations, which can be grouped into three main categories with different resolutions and box sizes ($\sim$50, 100, and 300~Mpc). Each hydrodynamical simulation is run together with two lower-resolution versions of it, and a dark-matter only version accompanying each of them. The simulation outputs are in the form of 100 snapshots with redshifts from $z=127$ to $z=0$. Each snapshot provides information about dark matter particles, star particles, gas cells and black holes. Halo, subhalo (galaxy) catalogs, and merger trees are also provided for each simulation.

In this paper, we mainly use the redshift $z=0$ snapshot of the TNG300-1 simulation (TNG300 hereafter), which has a box size of approximately 300~Mpc in comoving units, and it has the highest resolution among the 300~Mpc category. TNG300 has a dark matter particle mass of $3.98\times10^7 \msun$/h, an average gas cell mass of $7.44\times10^6 \msun$/h. The average mass of star particles is $7.8\times 10^{6}\msun$/h. The resolution of galaxies, calculated as approximately 100 times the average star particle mass, is $\sim10^{9}\msun$. Compared with TNG300, the resolution of TNG100 is approximately an order of magnitude higher. The large box size of TNG300 helps us to achieve better signals, especially for massive galaxies. However, as it has been discussed in previous studies, properties of simulated galaxies may not be necessarily fully converged in TNG300, though it explores larger volume \citep[e.g.][]{2018MNRAS.475..648P,2017MNRAS.465.3291W,2018MNRAS.473.4077P}. Thus in order to make sure that our results are not limited by resolution, we also compare the predictions by TNG100-1 (TNG100 hereafter). We do not show TNG100 results repeatedly. This is because due to the smaller volume, TNG100 results are noisier. However, we do have double checked that the conclusions in this paper do not show prominent differences across TNG300 and TNG100.

\subsection{Isolated central galaxies in observation and TNG}
\label{sec:icg}

In real observation, it is difficult to directly identify galaxies which are true central galaxies of dark matter halos. In order to select a sample with a high fraction of true halo central galaxies, we follow the method of \citet{2021ApJ...919...25W} to select those so-called isolated central galaxies, which are the brightest within a given volume. This is because the actual central galaxies of dark matter halos are often brighter than other companion galaxies around them. The sample of isolated central galaxies (hereafter ICGs or ICG) are selected from the spectroscopic Main galaxies of the seventh data release of the Sloan Digital Sky Survey \citep[SDSS/DR7;][]{2009ApJS..182..543A}. The parent sample used for selection is from the New York University Value Added Galaxy \citep[NYU-VAGC;][]{2005AJ....129.2562B} catalog, which is flux limited down to $r\sim17.7$.

An ICG is defined as the brightest galaxy in $r$-band magnitude, within a cylinder in redshift space, formed by a circular region in projected sky with a radius of virial radius and the height of the cylinder being $\pm$3 times the virial velocity along the line of sight. In addition, the ICG should not be within such a volume of another more massive galaxy in stellar mass. Here the virial radius and velocity are estimated through the abundance matching formula of \citet{2010MNRAS.404.1111G}. Throughout this paper, our measurements of the lensing signals, the projected stellar mass density profiles for these ICGs $+$ their extended stellar halos, and for surrounding satellites will all be centered on this sample of ICGs. 

To ensure that no ICG candidates have brighter companions without spectroscopic redshifts, we have used the photoz probability distribution catalog of \citet{2009MNRAS.396.2379C} to look for additional companions missed in the spectroscopic sample. We further eliminate any candidate with a companion in this catalog of equal or brighter $r$-band magnitude and projected within the virial radius, unless the photometric redshift distribution of the potential companion is inconsistent\footnote{The photoz probability distribution of the companion gives a less than 10\% of probability that it shares the same redshift as the central galaxy.} with the spectroscopic redshift of the candidate. We refer to \citet{2021ApJ...919...25W} for further details on the galaxy selection and completeness of this sample.

To select ICGs in TNG, we follow exactly the same isolation criteria as for real observation. Here we choose the $z$-axis as the line of sight, and the $z$ coordinate of each galaxy is displaced according to its velocity along the $z$-axis, to mimic the redshift space distortion of real galaxies. The virial radius and velocity are estimated in the same way as for real observed ICGs, instead of directly using the values from the simulation. Notably, the magnitudes used to select simulated ICGs are those which involved galaxy dust attenuation and solar neighborhood extinctions \citep{2018MNRAS.475..624N}. \cite{2018MNRAS.475..624N} claims significant improvements with respect to the original Illustris simulations after they apply this new dust model (model C in their paper). With the sample of ICGs selected from TNG300, we can directly check the fraction of true halo central galaxies. We find the fraction is as high as 89.1\%, with 10.9\% of satellite contamination.

In our analysis, we will separate our sample of observed and simulated ICGs into red and blue populations. The separation in real observation is drawn according to the $g-r$ color-mass diagram of SDSS spectroscopic Main galaxies. For TNG, the separation is determined according to the $g-r$ color-mass diagram of TNG galaxies, i.e., different from observation. For TNG galaxies, the colors have included the dust model \citep{2018MNRAS.475..624N}, which directly affects the color separation. 
The explicit color cuts are $g-r=0.065\log_{10}M_\ast/\msun+0.1$ for observation and $g-r=0.03\log_{10}M_\ast/\msun+0.4$ for TNG300. Table~\ref{tab:ngal_table} shows the number of red and blue ICGs in real observation and in TNG300, divided into different stellar mass bins.

In this paper, when we try to validate our lensing signal measurements, we will compare our measured signals with the previous measurements in \cite{2016MNRAS.457.3200M} and \citet{2016MNRAS.456.2301W}. \cite{2016MNRAS.457.3200M} performed their measurements by using a sample of so-called locally brightest galaxies (LBGs). The sample of LBGs are selected in a similar way as our sample of ICGs, but with different selection criteria. Explicitly, LBGs are the brightest in $r$-band within 1~Mpc in projection and $\pm$1000~km/s along the line of sight. We refer to \citet{2016MNRAS.456.2301W} for details. This sample of LBGs has been used for stacking weak lensing, X-ray and SZ signals \citep[e.g.][]{2013A&A...557A..52P,2015MNRAS.449.3806A,2015PhRvL.115s1301H,2016A&A...586A.140P,2016MNRAS.457.3200M} over a wide range in stellar mass, and historically, they were referred to as LBGs, so we maintain this nomenclature to distinguish them from our sample of ICGs. In this paper, the sample of LBGs will only be used when we compare our lensing signals with previous studies and when we test the effect of stellar mass corrections (Section~\ref{section:masscorr}).

\begin{table}[h!]
  \begin{center}
    \begin{tabular}{|c|c|cc|cc|}
      \hline
      \multicolumn{1}{|c|}{Stellar mass}
      &\multicolumn{1}{c|}{$R_{200}$}
      &\multicolumn{2}{c|}{SDSS}
      &\multicolumn{2}{c|}{TNG300}\\\cline{3-6}
      \multicolumn{1}{|c|}{[$\log_{10}M_\ast/\msun$]}&
      \multicolumn{1}{c|}{[kpc]}&
      \multicolumn{1}{c}{red}&
      \multicolumn{1}{c|}{blue}&
      \multicolumn{1}{c}{red}&
      \multicolumn{1}{c|}{blue}\\
      \hline
      11.4-11.7 & 758.00 & 1723 & 22,840 & 79 & 852  \\
      11.1-11.4 & 459.08 & 9056 & 42,006 & 172 & 2127   \\
      10.8-11.1 & 288.16 & 33,010 & 78,461 & 970 & 4741   \\
      10.5-10.8 & 214.80 & 50,499 & 61,813 & 5100 & 7984   \\
      10.2-10.5 & 173.18 & 43,011 & 31,183 & 13,137 & 3673  \\
      \hline
    \end{tabular}
    \caption{The virial radius, $R_{200}$, of ICGs in different stellar mass bins, and the number of red and blue ICGs in observation and TNG300. Here $R_{200}$ is defined as the radius within which the matter density is 200 times the critical density of the universe. $R_{200}$ values are taken from \cite{2021ApJ...919...25W} and are based on ICGs selected from a mock galaxy sample of the Munich semi-analytical model. Note that we use the same $R_{200}$ for red and blue ICGs in a given stellar mass bin.}
    \label{tab:ngal_table}
  \end{center}
\end{table}

\subsection{Weak Lensing Data}
\label{sec:sheardata}
The Dark Energy Spectroscopic Instrument \citep[DESI;][]{2016arXiv161100036D} is a spectroscopic project that aims to obtain optical spectra for millions of galaxies, quasars and Milky Way stars over a 5 year period of time. The DESI Legacy Imaging Surveys \citep{2019AJ....157..168D} are the precedented programs to take source images and prepare input photometric target sources for the DESI spectroscopic observations. It includes three imaging projects (the Dark Energy Camera Legacy Survey, the Beijing–Arizona Sky Survey, and the Mayall $z$-band Legacy Survey), and cover a total of $\approx14,000$ square degrees of the sky in three optical bands ($g$, $r$, and $z$). Among the three imaging surveys, the Dark Energy Camera Legacy Survey covers the North Galactic Cap region of the sky at $\mathrm{Dec}\leq32^\circ$ and the South Galactic Cap region of the sky at $\mathrm{Dec}\leq34^\circ$, reaching $5\sigma$ depths of 24.0, 23.4, and 22.5 in magnitudes for the $g$, $r$ and $z$ bands, respectively.

In this study, our weak lensing shear data is constructed from the Dark Energy Camera Legacy Survey (DECaLS) DR8 data using \textsc{Fourier\_Quad} (see Section~\ref{section:FQ} for details). 
The shear catalog itself has passed the field distortion test\footnote{The field distortion is an optical aberration found in all optical systems that produces a distortion in the galaxy shapes similar to the cosmic shear.} \citep{2019ApJ...875...48Z}. According to \cite{2022AJ....164..128Z}, the multiplicative biases in DECaLS $r$ and $z$-bands are all in reasonable ranges. However, in DECaLS DR8, the $g$-band data shows a larger bias compared to $r$ and $z$ bands in the initial field distortion test. Therefore, we only use the $r$ and $z$-bands data in this work.

6Our galaxy catalog for the above shear measurements is provided by \citet{2019ApJS..242....8Z}. \citet{2019ApJS..242....8Z} presents a catalog of approximately 0.19 billion galaxies with $r<23$~mag in DECaLS, in three optical bands ($g$, $r$, and $z$), and two infrared ones. We use the photometric redshifts provided by \citet{2021MNRAS.501.3309Z}, which adopts a machine learning approach to compute photometric redshifts. Concretely, they applied a random forest regression model, a popular supervised learning algorithm composed by several decision trees. For the training process, they combined the data from 10 spectroscopic surveys to form a training set of what they called "truth" redshift values. 

\subsection{The Hyper Suprime-Cam Survey imaging products and photometric sources}
\label{sec:hsc}

The Hyper Suprime-Cam Subaru Strategic Program (HSC-SSP) project \citep{2018PASJ...70S...4A,2018PASJ...70S...1M,2012SPIE.8446E..0ZM,2018PASJ...70S...2K,2018PASJ...70S...3F} is a wide-field multi-band (\textit{g,r,i,z,y}) imaging survey that uses the Hyper Suprime-Cam (HSC) on the Subaru $8.2m$ telescope. The project works on three levels of depth: Wide, Deep, and Ultradeep, which aims to cover a total of $1,400$, $27$, and $3.5$ square degrees of footprints, and with $r=26$, $r=27$, and $r=28$ depths in magnitudes, respectively. In this paper, we use photometric galaxies from the wide area of the third public data release (pDR3) to measure the projected stellar mass density profiles of satellite galaxies around our sample of ICGs. pDR3 \citep{2022PASJ...74..247A} covers 670 square degrees in all five bands, in the Wide layer. Besides, the HSC coadded imaging product was used by \cite{2019MNRAS.487.1580W} and \cite{2021ApJ...919...25W} to measure the surface brightness and projected stellar mass density profiles of ICGs and their stellar halos, and the measurements of \cite{2021ApJ...919...25W} will be compared with TNG predictions in this study. In the following, we briefly introduce the HSC imaging products and photometric sources. 

HSC-SSP data is processed with the HSC pipeline, which is a specialized version of the LSST \citep{2010SPIE.7740E..15A,2017ASPC..512..279J} pipeline code. HSC ``visited'' the same sky for multiple times. The HSC pipeline involves four main steps: (1) processing of single exposure (one visit) image (2) joint astrometric and photometric calibration (3) image coaddition and (4) coadd measurement. In the first step, bias, flat field and dark flow are corrected for. Bad pixels, pixels hit by cosmic rays and saturated pixels are masked and interpolated. The pipeline first performs an initial background subtraction before source detection. Detected sources are matched to external reference catalogs in order to calibrate the zero point and a gnomonic world coordinate system for each CCD. After galaxies and blended objects are filtered out, the sky background is estimated and subtracted again. A secure sample of stars are used to construct the PSF model. In the second joint calibration step, the astrometric and photometric calibrations are refined by requiring that the same source appearing on different locations of the focal plane during different visits should give consistent positions and fluxes \citep{2018PASJ...70S...5B}. In the third step, the HSC pipeline resamples images to the predefined output skymap. It involves resampling both the single exposure images and the PSF model \citep{2011PASP..123..596J}. Resampled images of different visits are then combined together (coaddition). Images produced through coaddition are called coadded images.

The last step is the most relevant for the measurement of the projected stellar mass density profiles of photometric satellites. In this step, photometric sources are detected, deblended and measured from the coadded images. A maximum-likelihood detection algorithm is run independently for each band at first. Background is estimated and subtracted once again. The detected footprints and peaks of sources are merged across different bands to maintain detections that are consistent over bands and to remove spurious peaks. These detected peaks are deblended and a full suite of source measurement algorithm is run on all objects, yielding independent measurements of source positions and other properties in each band. A reference band of detection is then defined for each object based on both the signal-to-noise ratio (SNR) and the purpose of maximizing the number of objects in the reference band. Finally, the measurement of sources are run again with the position and shape parameters fixed to the values in the reference band to achieve the ``forced'' measurement. The forced measurement brings consistency across bands. 

For analysis in this paper, we select those photometric sources\footnote{The sources are used to calculate projected stellar mass density profiles of photometric satellites, so different from background photometric sources used to calculate lensing signals from DECaLS.} which are classified as extended, i.e., galaxies. We only use galaxies which are brighter than 25 in $r$-band. As has been discussed by \cite{2018PASJ...70S...8A} and \cite{2021MNRAS.500.3776W}, the completeness of photometric sources in HSC is very close to 1 at $r\sim25$. We also exclude sources with any of the following flags set as true in $g$, $r$ and $i$-bands: bad, crcenter, saturated, edge, interpolatedcenter or suspectcenter. Note, in principle, we can also use the photometric sources from DECaLS, i.e., the imaging survey adopted for our weak lensing shear measurements, to measure the projected stellar mass density profiles of photometric satellites. However, as having been shown by \cite{2021MNRAS.500.3776W}, DECaLS and HSC photometric sources give fully consistent results, and although HSC is much smaller in footprint, the signals measured from HSC are comparable to or sometimes slightly better than DECaLS, due to its much deeper flux limit. Besides, the projected stellar mass density profiles of ICGs and their stellar halos, as measured by \cite{2021ApJ...919...25W} and will be investigated in this study, are also based on HSC galaxy images. Thus we choose to use HSC photometric sources when we investigate satellite galaxies in this study.

\section{Methods} \label{section:Methods}

\subsection{The \textsc{\textsc{Fourier\_Quad}} Method and Lensing Signal Measurements}

\subsubsection{\textsc{Fourier\_Quad}}
\label{section:FQ}

\textsc{Fourier\_Quad} \citep[]{2015JCAP...01..024Z,2017ApJ...834....8Z,2019ApJ...875...48Z} is a unique shear measurement pipeline that measures the shear signal in Fourier space, based on the weighted moments of the power spectrum of the galaxy images. Here we briefly describe \textsc{Fourier\_Quad} and the readers can refer to \citet{2019ApJ...875...48Z} for more details.

\textsc{Fourier\_Quad} calculates the 2D power spectrum of the sources. 
The shear estimators are defined as:
\begin{equation}
G_1 = -\frac{1}{2}\int{d^2\vec{k}(k^2_x-k^2_y)T(\vec{k})M(\vec{k})},
\label{eq_shear_1}
\end{equation}
\begin{equation}
G_2 = -\int{d^2\vec{k}k_x k_y T(\vec{k})M(\vec{k})},
\label{eq_shear_2}
\end{equation}
\begin{equation}
N = \int{d^2\vec{k}\left[k^2-\frac{\beta^2}{2}k^4\right]T(\vec{k})M(\vec{k})},
\end{equation}

where $\vec{k}$ is the wave vector, and
\begin{equation}
T(\vec{k}) = \left|\tilde{W}_{\beta}(\vec{k}) \right|^2 /  \left|\tilde{W}_{PSF}(\vec{k}) \right|^2,
\end{equation}
\begin{equation}
M(\vec{k}) = \left|\tilde{f}^S(\vec{k}) \right|^2 - \tilde{F}^S -  \left|\tilde{f}^B(\vec{k}) \right|^2 + \tilde{F}^B.
\end{equation}

$T(\vec{k})$ represents the ratio between the power spectrum of an isotropic Gaussian function, $\tilde{W}_{\beta}(\vec{k})$, and the power spectrum of the original PSF. $\tilde{W}_{\beta}(\vec{k})$ is defined as:
\begin{equation}
    \tilde{W}_{\beta}(\vec{x}) = \frac{1}{2\pi \beta^2} \mathrm{exp}\left(-\frac{\left|\vec{x} \right|^2}{2\beta^2} \right),
\end{equation}
where $\beta$ is the scale radius of the Gaussian function.
$T(\vec{k})$ transforms the form of the original PSF to the desired Gaussian form in order to correct for the PSF effect. \citet{2015JCAP...01..024Z} claims the ideal value of $\beta$ to be slightly larger ($20-50\%$) than the scale radius of the original PSF, to avoid singularities in the conversion. Additionally, the value of $\beta$ should not be too large, in order to avoid information loss. $M(\vec{k})$ corrects the power spectrum of the source by subtracting the contributions of the background and the Poisson noise. $\tilde{f}^S(\vec{k})$ and $\tilde{f}^B(\vec{k})$ are the Fourier transformations of the galaxy and the background noise, respectively. $\tilde{F}^\mathrm{S}$ and $\tilde{F}^\mathrm{B}$ are estimates of the power spectrum of the Poisson noise for the galaxy image and the background, respectively.

It can be shown that the ensemble averages of the shear estimators so defined satisfy the following relations \citep{2011MNRAS.414.1047Z} :
\begin{equation}
    \frac{\left<G_1 \right>}{\left<N \right>}=g_1+O(g_{1,2}^3), \frac{\left<G_2 \right>}{\left<N \right>}=g_2+O(g_{1,2}^3),
\end{equation}
where $g_1$ and $g_2$ are the two shear components. In a more recent work, \citet{2017ApJ...834....8Z} proposed a new approach for the shear measurement, called the PDF Symmetrization Method (PDF-SYM). PDF-SYM symmetrizes the probability distribution function (PDF) of $G_1-\hat{g}_1\left(N+U \right)$ and $G_2-\hat{g}_2\left(N-U \right)$ in order to determine the shear values $\hat{g}_1$ and $\hat{g}_2$. This introduces two additional terms ($U$ and $V$) to the three initially defined ($G_1$, $G_2$, and $N$). The new terms are defined as:

\begin{equation}
\label{Ud}
U = -\frac{\beta^2}{2}\int{d^2\vec{k} (k_x^4-6k_x^2k_y^2+k_y^4)T(\vec{k})M(\vec{k})},
\end{equation}
\begin{equation}
\label{Vd}
V = -2\beta^2 \int{d^2\vec{k} (k_x^3k_y-k_xk_y^3)T(\vec{k})M(\vec{k})}.
\end{equation}
$U$ involves the parity properties of the shear estimators, whereas $V$ is needed to transform $U$ when a coordinate rotation appears in the shear measurement. \citet{2017ApJ...834....8Z} proved that PDF-SYM allowed the shear estimation to reach the lowest theoretical statistical limit (Cramer-Rao limit). 

\textsc{Fourier\_Quad} has been validated by the field distortion test (see Section~\ref{sec:sheardata} above), and the readers can find more details in \cite{2019ApJ...875...48Z} and \cite{2022AJ....164..128Z}.

\subsubsection{The observed lensing signals}
\label{sec:lensobsmethod}

In this section we describe the method used for the lensing signal calculation in real observation. In weak lensing, the tangential component of the shear signal $\gamma_\mathrm{t}$ is related to the density profile of the foreground lens via:
\begin{equation}
\Delta\Sigma(r_\mathrm{p}) = \gamma_\mathrm{t} \Sigma_{\mathrm{cr}},
\label{eq:eq_sigma_2_1}
\end{equation}
in which $\Delta\Sigma$ is the excess surface density defined as:

\begin{equation}
\Delta\Sigma(r_\mathrm{p}) = \bar{\Sigma}(<r_\mathrm{p}) - \Sigma(r_\mathrm{p}), 
\label{eq:eq_sigma_1}
\end{equation}
where $\bar{\Sigma}(<r_\mathrm{p})$ is the averaged projected mass density within radius $r_\mathrm{p}$, and $\Sigma(r_\mathrm{p})$ is the projected mass density or surface density at radius $r_\mathrm{p}$.
$\Sigma_{\mathrm{cr}}$ is the critical surface density, defined as
\begin{equation}
\Sigma_{\mathrm{cr}} = \frac{c^2}{4\pi G} \frac{d_\mathrm{s}}{d_\mathrm{l} d_{\mathrm{ls}} },
\label{eq:eq_sigma_2_2}
\end{equation}
in physical units. Here $d_\mathrm{s}$, $d_\mathrm{l}$, and $d_\mathrm{ls}$ are the angular diameter distances to the source, to the lens, and between source and lens, respectively.

Conventionally, the lensing signal is calculated as the weight sum of all source-lens pairs. In this project, we follow a different approach, based on the PDF-SYM method. As described in section~\ref{section:FQ}, \textsc{Fourier\_Quad} derives the shear as the value that better symmetrizes the PDF of the shear estimators. For the lensing signals, this means that we need to symmetrize the following PDF,
\begin{equation}
\label{Gt}
G_\mathrm{t}-\frac{\widehat{\Delta\Sigma}}{\Sigma_\mathrm{cr}(z_\mathrm{l},z_\mathrm{s})}\left(N+U_t \right),
\end{equation}
in which $G_\mathrm{t}$ is the tangential shear estimator, which is derived from $G_1$ and $G_2$ defined in Equation~\ref{eq_shear_1} and ~\ref{eq_shear_2} through a coordinate rotation. Similarly, $U_t$ is from rotating the the $U$ and $V$ components defined in Equation~\ref{Ud} and ~\ref{Vd}. $\widehat{\Delta\Sigma}$ is the excess surface density that one seeks to symmetrize the PDF of the tangential shear estimator defined in Equation~\ref{Gt}. We set a photometric redshift cut of $\Delta z=0.2$ between source and lens in order to mitigate the dilution effect of foreground galaxies.

A few recent works have used this shear measurement method and have calculated the lensing signals with it. For example, \cite{2022MNRAS.513.4754F} performed the first measurements of the so called characteristic depletion radius using weak lensing data. \cite{2022ApJ...936..161W} made use of the same weak lensing data to study halo properties of different lens samples.

In Appendix~\ref{Appendix:Appendix_A}, we show a comparison between our calculated lensing signals based on \textsc{Fourier\_Quad} shear estimator and the previous measurement of \cite{2016MNRAS.457.3200M} and \cite{2016MNRAS.456.2301W} based on shear measurements of SDSS. The signals are centered on the same sample of LBGs. The agreement is very good, which together with the excellent results on the field distortion test \citep{2019ApJ...875...48Z,2022AJ....164..128Z}, indicating that we can safely use the lensing results based on the \textsc{Fourier\_Quad} method for our analysis throughout this paper.

\subsubsection{Lensing signals in TNG}

For TNG, the excess surface density can be directly calculated based on the spatial distribution of all particles, following Equation~\ref{eq:eq_sigma_1}. We include dark matter particles, stars, wind particles, gas cells, and black holes. Radial bins are defined in projected distance, $r_\mathrm{p}$. ${\bar{\Sigma}(<r_\mathrm{p})}$ includes all particles within radius $r_\mathrm{p}$, and $\Sigma(r_\mathrm{p})$ includes all particles between $r_\mathrm{p}-\Delta r_\mathrm{p}$ and $r_\mathrm{p}+\Delta r_\mathrm{p}$, where $\Delta r_\mathrm{p} << r_\mathrm{p}$.

ICGs in real observation are flux limited, whereas ICGs in TNG are volume limited. In order to consider the flux limit in our observational data and ensure fair comparison between observation and TNG300, we weight each ICG in TNG by a maximum volume when calculating the excess surface density profiles. The maximum volume is calculated from the limiting redshift or distance when one can still observe the galaxy given the SDSS flux limit, by placing simulated galaxies from the $z=0$ snapshot at a redshift where the apparent magnitudes
reach $r=17.7$ (flux limit of our ICGs from the SDSS Main galaxy sample). 
Similar weights are assigned when we compute the projected stellar mass density profiles for ICGs $+$ their stellar halos and for satellites (see subsections below). However, we find including or not including this weight show very small differences in our results, because ICGs are divided into a few narrow stellar mass bins in our results, and the flux limit barely affects the median or mean stellar mass and halo mass given the narrow bin width.

\subsection{Projected Stellar Mass Density Profile of satellites}
\label{section:proj_SM}

\subsubsection{Observed satellite profiles}
In real observation, we follow the approach of \cite{2021ApJ...919...25W} to measure the projected stellar mass density profiles of satellites (hereafter satellite profiles). We adopt a satellite mass threshold of $M_\ast>10^9\msun$, that is, only satellites more massive than this threshold can contribute to the final signals. The threshold is chosen to match the resolution limit for TNG to ensure fair comparisons. The readers can refer to \cite{2021ApJ...919...25W} and \cite{2012MNRAS.424.2574W} for details, and here we briefly introduce the method.

The observed satellite profiles are measured by counting photometric galaxy companions around spectroscopically identified ICGs. The photometric galaxies are from HSC pDR3 (see Section~\ref{sec:hsc}). In our method, we first bin ICGs according to their stellar masses or colors, and for a given bin, we loop through all ICGs. For each ICG, we first assume that all its companions are at the same redshift, and we use the spectroscopic redshift of this ICG to calculate the intrinsic luminosity and rest frame colors of the companions. Here we adopt the $K$-correction formulas of \cite{2010PASP..122.1258W}. The stellar mass of each companion is further calculated through an empirical relation between the stellar-mass-to-light ratio ($M/L$) and the rest-frame color (see \cite{2012MNRAS.424.2574W} for details). The cumulated stellar masses of satellites as a function of projected radius for ICGs in the same bin are added together and averaged in the end.  

We then correct the contamination by foreground and background sources, we repeat exactly the same steps around a sample of random points within the HSC footprint. The random points are assigned the same redshift and stellar mass distributions as real ICGs. The averaged companion profiles around the random points in different stellar mass and color bins are statistical estimates of the foreground $+$ background source contamination, and are subtracted off from the profiles around real ICGs. 

The photometric sources of HSC are approximately fluxed limited down to 26 in $r$-band apparent magnitude, and we adopt a safe flux limit of $r=25$. To ensure the completeness of our satellite counts at $M_\ast>10^9\msun$, the flux limit is transferred to a lower limit in stellar mass, again based on the redshift of the central ICG and the $M/L$ estimated from the reddest color allowed at the corresponding redshift. Here the reddest color is calculated through the red end boundary of all spectroscopic SDSS galaxies at this redshift, which corresponds to the largest allowed value of $M/L$, and thus the safest lower limit in stellar mass. Only when this mass limit is smaller than $M_\ast=10^9\msun$, companions around this ICG will be cumulated and averaged in the end. 

Notably, \cite{2021ApJ...919...25W} showed that observed satellite profiles can be severely affected by deblending mistakes of the central dominant ICG within $0.1R_{200}$, where $R_{200}$ is the virial radius of the host dark matter halo within which the enclosed matter density is 200 times the critical density of the Universe\footnote{In this paper, our definition of the virial mass, $M_{200}$, is the total mass enclosed within $R_{200}$.}. Thus throughout this paper, when discussing satellite profiles, we should only focus on radial ranges beyond $0.1R_{200}$. 

The above methods have been robustly developed and carefully tested in \cite{2011ApJ...734...88W,2012MNRAS.424.2574W,2014MNRAS.442.1363W}. In fact, similar methodology has been independently developed and applied to many different modern surveys and varying types of galaxies at different redshifts in the last decade or so \citep[e.g.][]{2011AJ....142...13L,2011MNRAS.417..370G,2012ApJ...760...16J,2012MNRAS.427..428G,2014ApJ...792..103K,2016MNRAS.459.3998L,2022ApJ...925...31X,2022ApJ...926..130X,2022ApJ...939..104X,2022arXiv221102665X}. The very early idea can be traced back to \cite{1994MNRAS.269..696L}.

\subsubsection{Satellite profiles in TNG}

For TNG, in principle, we directly know from the simulation the true population of satellite galaxies bound to each ICG. However, in order to ensure fair comparisons with observation, we calculate satellite profiles in a way similar to real observation. We choose the $z$-axis of the simulation box as the line-of-sight direction, and count the number of companions around our selected sample of TNG ICGs in projected radial bins perpendicular to the artificial line of sight, i.e., the $x-y$ plane. We count all the companions projected around the ICG, i.e., not only those true satellites, but also those foreground and background galaxies. When cumulating the stellar mass as a function of projected radius to the central ICG, the stellar mass of each companion is the total stellar mass of all bound star particles belonging to it. Only companions more massive than $M_\ast=10^9\msun$ are used, which is approximately the resolution limit of satellite galaxies in TNG300-1. Similar to how we obtain the lensing signals, we weight each ICG by the maximum volume to which one can still observe this galaxy given the flux limit of SDSS. 

We then subtract the foreground and background contamination, which is calculated by randomly generating $10^4$ points within the simulation box and calculating their companion profiles in the same way as for ICGs. Moreover, since the simulation box is regular, we can also simply estimate the level of background contamination from the average volume density of galaxies in the simulation box and the volume of each projected radial bin. This gives nearly identical results as when the background contamination is estimated from companions around random points in the simulation box. 

Rigorously, in order to fully mimic real observation, we have to generate light-cone mock observations for simulated galaxies, so that we have realistic flux limits, background contamination and galaxy evolution. However, most of our ICGs are at $z<0.25$, with the most massive ICGs extending to at most $z\sim0.3$. The evolution of galaxies at $z<0.3$ is weak. Besides, as having been shown by \cite{2012MNRAS.424.2574W}, directly projecting the simulation box and light-cone mocks lead to very similar results in the final stacked/averaged surface brightness, with the noise in the former suppressed lower, and thus we choose to directly project the simulation box. 

\begin{figure*}
    \begin{center}
	\includegraphics[width=0.9\textwidth]{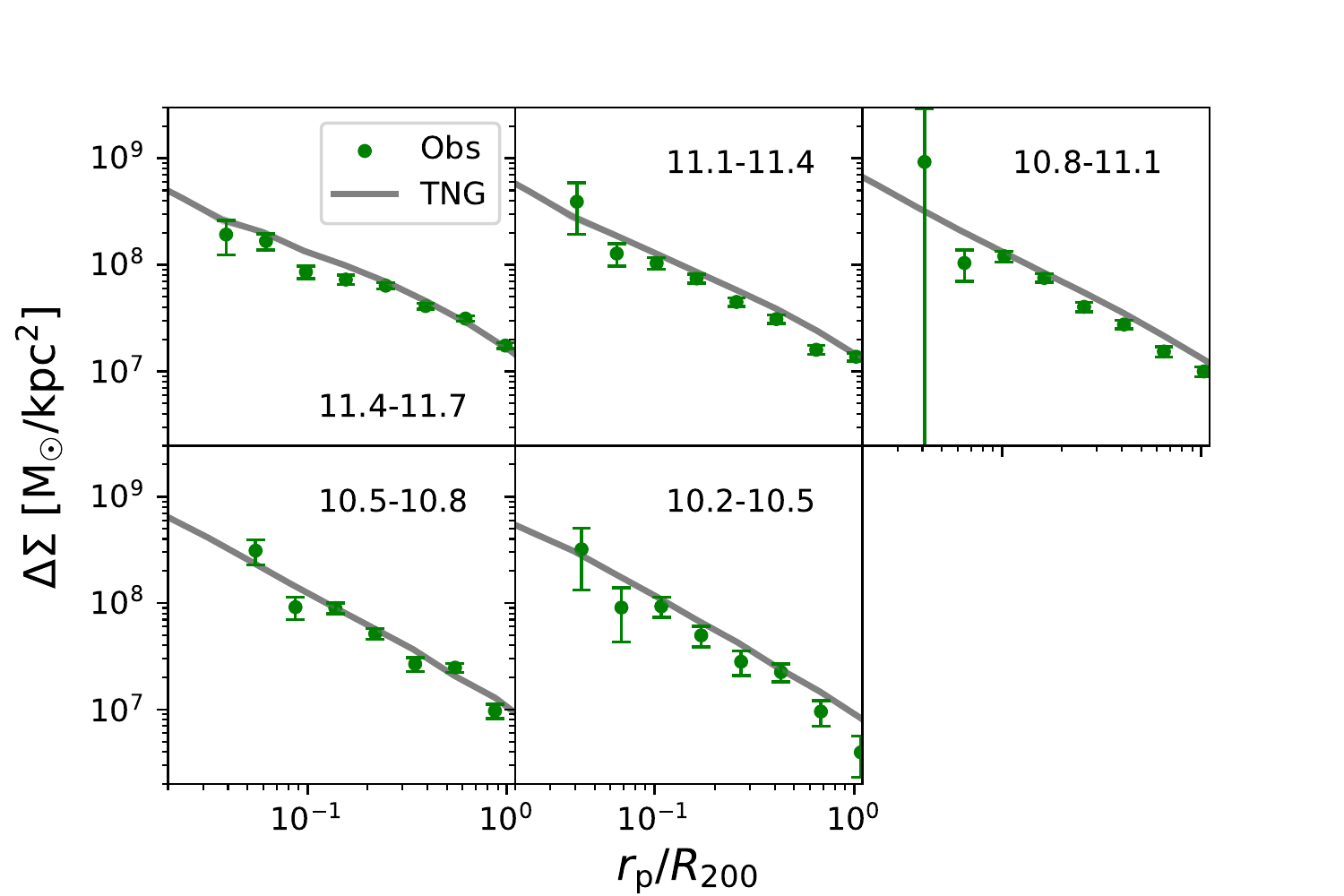}
	\caption{Lensing signals for ICGs in real observation (green points) and TNG300 (gray lines). Each panel corresponds to ICGs in a different stellar mass range, as indicated by the text, which is the stellar mass range in unit of $\log_{10}\msun$. Observed lensing signals are measured using \textsc{Fourier\_Quad}, with errorbars calculated as the 1-$\sigma$ scatter of 200 jackknife subsamples. The subsampling is applied to the lens catalog. Errors for TNG300 are calculated as the 1-$\sigma$ scatters of 100 bootstrap subsamples, which are compared to the line width. Stellar mass corrections have been included to account for the missing stellar mass in outskirts of massive galaxies in observation.} 
	\label{fig:figure_1}
	\end{center}
\end{figure*}

\subsection{Projected stellar mass density profiles for ICGs and their stellar halos in observation and TNG}
\label{section:proj_SM2}

\subsubsection{Observed central profiles}

In real observation, the projected stellar mass density profiles of ICGs and their stellar halos (hereafter central profiles) are directly taken from \cite{2021ApJ...919...25W}, and here we only briefly introduce the method. 

The ICG surface brightness profiles are obtained by stacking galaxy images from the HSC coadded images (see Section~\ref{sec:hsc} for details), which are then converted to projected stellar mass density profiles based on the PSF corrected $g-r$ color profiles \citep{2021ApJ...919...25W}. Explicitly, image cutouts are at first created for each ICG, with bad pixels masked, such as those hit by cosmic rays and saturated pixels. For each cutout, companion sources, including true satellite galaxies, fore/background sources are detected by Sextractor and masked. \cite{2021ApJ...919...25W} have tried different source detection and masking thresholds, to ensure safe masking of companion sources. Each cutout is resampled to exactly the same world coordinate system and the same output grids, which we call as input images. For each pixel in the output image plane, we at first clip corresponding pixels from all input images (masked pixels are not included). Say all input images, if not masked at this pixel, contribute $N$ reads in surface brightness. We discard 10\% pixels near the two distribution tails of all input pixel values, in order to avoid extreme pixel values biasing our results. After the clipping we take the average of all remaining pixel values. The choice of 10\% follows the value used by \cite{2014MNRAS.443.1433D} and is empirical. \cite{2021ApJ...919...25W} tested that choices of 1\% to 10\% lead to very similar results. In the end, we repeat exactly the same steps for the images around a sample of random points within the HSC footprint. The random stacks are estimates of the residual sky backgrounds, which are subtracted off from the stacks for real ICGs.

\subsubsection{Central profiles in TNG}
For TNG, central profiles are directly calculated by using all bound star particles belonging to each ICG and those diffuse star particles not belonging to any other galaxies but are in the same friends-of-friends (FoF) group. Again we choose the $z$-axis as the line-of-sight direction. Projected radial bins are arranged in the same way as for satellite profiles.

\subsection{Stellar mass correction}
\label{section:masscorr}

The stellar masses we use to divide ICGs into different stellar mass bins are initially from SDSS, which were measured by fitting the stellar population synthesis model \citep{2007AJ....133..734B} assuming a Chabrier (2003) initial mass function to the SDSS photometry. However, as have been pointed out in many previous studies \citep[e.g.][]{2005AJ....130.1535G,2007MNRAS.379..867V,2011ApJS..193...29A,2013MNRAS.436..697B,2013ApJ...773...37H,2015MNRAS.454.4027D}, the total stellar mass of massive central galaxies could be underestimated, due to the fact that their outskirts contribute a non-negligible amount fraction of the total stellar mass but are often oversubtracted as backgrounds or are under the detection threshold, especially for shallow surveys. This leads to disagreement with predictions by numerical simulations, in terms of the stellar mass functions at the massive end \citep[e.g.][]{2013ApJ...773...37H,2015MNRAS.454.4027D}. The sample of ICGs (also LBGs) used in our study are selected from the SDSS spectroscopic main galaxies, and thus we need to correct for the missing part in their total stellar mass. This correction is im,portant to ensure fair comparisons with TNG300, because for TNG300 galaxies, we use their total stellar mass from the database, without any aperture cuts. 

To achieve the corrections, we use the data provided by the authors of \cite{2013ApJ...773...37H}. Explicitly, \cite{2013ApJ...773...37H} reanalyzed the images of a large sample of red elliptical galaxies from SDSS \citep[e.g.][]{2012MNRAS.427..146H}, and the photometry of these galaxies down to different sky limits is compared with the original SDSS photometry, which can be used as the reference for our stellar mass correction. We use the difference between their remeasured magnitudes and the original SDSS Petrosian magnitudes ($\Delta m_r$) for the correction. We convert $\Delta m_r$ to the corresponding correction in log stellar mass ($\Delta \log_{10}M_\ast$), simply through Equation~\ref{eq:eq_sm_correction}. 

\begin{equation}
\log_{10}M_\mathrm{new} = \log_{10}M_\mathrm{old} + \frac{\Delta m_r}{2.5}.
\label{eq:eq_sm_correction}
\end{equation}

For each massive ICG or LBG in our analysis, we match it to the closest one in $r$-band absolute magnitude from the data of \citet{2013ApJ...773...37H}, and obtain the newly corrected stellar mass according to Equation~\ref{eq:eq_sm_correction}. We only apply the correction for galaxies of $r$-band absolute magnitude brighter than -22.5, since smaller galaxies are expected to suffer less from such underestimates in their stellar mass and \cite{2013ApJ...773...37H} only considers galaxies in this magnitude range. 

A comparison of the lensing signals before and after the correction in stellar mass, is provided in Appendix~\ref{Appendix:Appendix_B}. For the two most massive bins in stellar mass, the difference before and after the correction is prominent. Without the correction, the values of stellar mass are smaller than the true stellar mass, and thus given a stellar mass bin, we are actually selecting more massive galaxies to fit into this bin. Moreover, at the massive end, a small increase in stellar mass would correspond to a more significant increase in host halo mass. As a result, the lensing signals are significantly higher in amplitudes in the two most massive bins. After the stellar mass correction, the amplitudes are significantly lower, bringing in good agreement with the prediction by IllustrisTNG300. Our results thus show that proper corrections to the stellar mass of massive galaxies in observation are crucial for fair comparisons with theoretical predictions. 

In our analysis, we apply the stellar mass correction above when calculating the lensing signals and the satellite profiles. The central profiles are directly taken from \cite{2021ApJ...919...25W} without including such corrections in stellar mass. More explicitly, the lensing and satellite profiles are binned by stellar masses after the correction. The central profiles are binned according to the original stellar mass without such corrections. However, we will show that our conclusions based on the central profiles in this paper are not affected by the stellar mass correction. 

\begin{figure*}
    \begin{center}
	\includegraphics[width=0.9\textwidth]{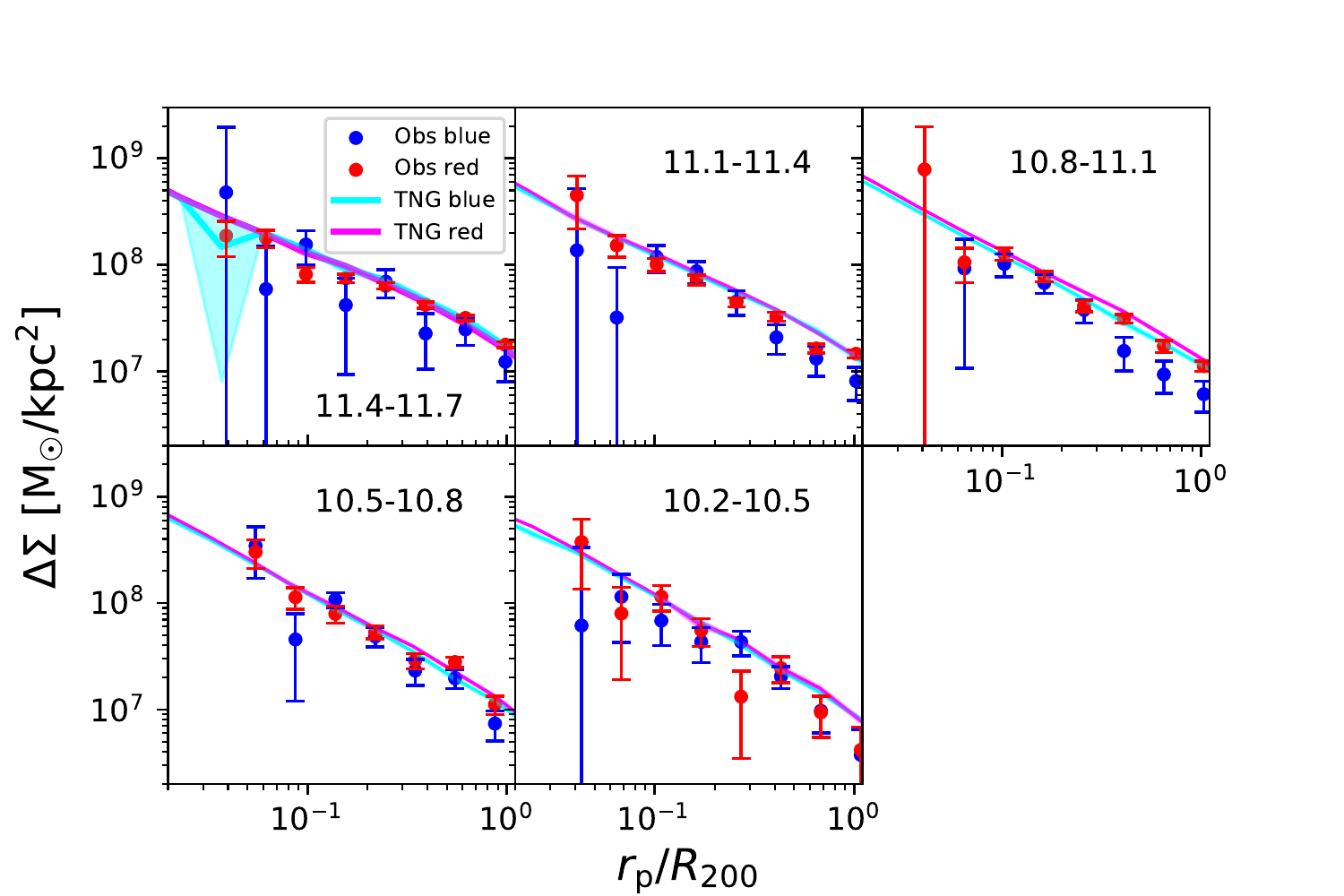}
	\caption{Lensing signals for observed blue (blue dots) and red (red dots) ICGs, and TNG300 blue (cyan curves) and red (magenta curves) ICGs. Each panel corresponds to a different bin in stellar mass, with the text in each panel showing the stellar mass range in unit of $\log_{10}\msun$. Errorbars for real observation are calculated as the 1-$\sigma$ scatters of 200 jackknife subsamples applied on the lens catalog. Errors for TNG300 are calculated as the 1-$\sigma$ scatters of 100 bootstrap subsamples, which are mostly comparable to the line width. Stellar mass corrections have been included to account for the missing stellar mass in outskirts of massive galaxies in observation.}
	\label{fig:figure_lensing}
	\end{center}
\end{figure*}

\section{Results} 
\label{section:Results}

In this section, we present our measured lensing signals, satellite and central  profiles around our sample of ICGs. The measurements are shown in parallel for both the real observation and predictions by TNG300, for direct comparisons. We have also confirmed that the trends are similar though noisier in TNG100. Part of the TNG100 predictions will be shown in Appendix~\ref{Appendix:Appendix_C}. In the end, we investigate the fraction of total stellar mass and the fraction of stellar mass in surviving satellites as a function of projected radius.

\subsection{Lensing signals}
\label{section:Results_2}

Figure~\ref{fig:figure_1} shows the lensing signals for real observation and TNG300, centered on ICGs grouped in five different stellar mass ranges. We scale the projected distance, $r_\mathrm{p}$, by the virial radius, $R_{\mathrm{200}}$, with the values of $R_{\mathrm{200}}$ for ICGs in different stellar mass bins taken from \cite{2021ApJ...919...25W}, which are based on ICGs selected from a mock ICG sample of the Munich semi-analytical model\footnote{As we have checked, the $R_{\mathrm{200}}$ values for different stellar mass bins based on the \cite{2011MNRAS.413..101G} model and based on the TNG simulations can be slightly different from each other, which may also differ from the actual $R_{\mathrm{200}}$ values of real galaxies, though in observation $R_{\mathrm{200}}$ cannot be measured directly. However, this does not affect our comparisons, because we scale observational results and TNG predictions by the same $R_{\mathrm{200}}$.}\ \footnote{In Section~\ref{sec:icg}, we use the virial radius predicted by the abundance matching formula of \citet{2010MNRAS.404.1111G} as the criterion to select ICGs. For the selection, each galaxy can have different virial radius based on their stellar mass. From now on, the virial radius, $R_{200}$, used to scale $r_\mathrm{p}$ is from mock ICGs of the \cite{2011MNRAS.413..101G} model, so not exactly the same as those calculated from the abundance matching formula of \citet{2010MNRAS.404.1111G}. We make such choices in order to be consistent with \cite{2021MNRAS.500.3776W}. For all ICGs in the same stellar mass bin, the $R_{200}$ value adopted to scale $r_p$ is the same, i.e., the mean $R_{200}$ for all mock ICGs in the bin. Also note although the explicit choice of $R_{200}$ can affect our ICG sample selection, this does not affect our results, because we select ICGs in TNG and real observation in exactly the same way, i.e., both are based on the same \citet{2010MNRAS.404.1111G} abundance matching $R_{200}$.}\citep[][]{2011MNRAS.413..101G} (see \cite{2012MNRAS.424.2574W} for details). We scale $r_\mathrm{p}$ by $R_{200}$ in order to investigate the density profiles over halo scales for different mass bins, and this will be the convention when we report our measurements in this paper. The $R_{200}$ values used to scale $r_\mathrm{p}$ are provided in Table~\ref{tab:ngal_table} for each stellar mass bin. 

We find good agreements between observed and TNG lensing signals. For the observational lensing signal calculation, a correction in stellar mass has been applied to massive galaxies (See Section~\ref{section:masscorr} and Appendix~\ref{Appendix:Appendix_B} for details). This correction significantly affects the signals in the two most massive bins, which improves the agreement between the real observation and TNG300 at the massive end. Without this correction in stellar mass, the observed lensing profiles are significantly higher than the simulation predictions in the two most massive panels (See figure~\ref{fig:lgbs_after_corr}). 

In a previous study, \cite{2020MNRAS.498.5804R} calculated the lensing signals based on the Munich semi-analytical model and TNG300. Good agreements were reported between the real signals and simulation predictions. \cite{2016MNRAS.456.2301W} reached similar results when comparing predictions by different Munich semi-analytical models \citep[e.g.][]{2011MNRAS.413..101G,2015MNRAS.451.2663H}. However, the simulation predictions are lower than the observed lensing signals for the most massive bin.  We think this is mainly due to the missing stellar mass in outskirts of our ICGs selected from SDSS, which were not properly corrected in \cite{2016MNRAS.456.2301W}.

In Figure~\ref{fig:figure_lensing}, we present our lensing signals for red and blue ICGs separately, and for both real observation and TNG300. Note as having been mentioned in Section~\ref{sec:icg}, we determine red and blue galaxies separately for real observation and TNG, based on the $g-r$ color-mass diagrams. The division slightly depends on stellar mass. In general, we find an overall good agreement between both lensing signals and for red and blue ICGs. However, in the outskirts of the three most massive panels, TNG predicted that signals around blue ICGs are higher in amplitudes than those around real observed blue ICGs. 

In previous studies based on numerical simulations or based on direct weak lensing measurements, it was reported that red galaxies are hosted by more massive dark matter halos than blue ones at fixed stellar mass \citep[e.g.][]{2012ApJ...757....4P,2012MNRAS.424.2574W,2016MNRAS.457.3200M,2019ApJ...881...74M}. This is also seen in our observed lensing signals. For example, in the $10.8<\log_{10}<11.1$ panel, the red dots are above the blue ones by $\sim3\sigma$ beyond $0.3R_{200}$. In the other two more massive panels there are also indications that the red dots are above blue ones at $r_\mathrm{p}>0.3R_{200}$, but with a lower significance with respect to the errors ($\sim1.8$ and $2\sigma$ between $0.3R_{200}$ and $R_{200}$ for the most and second massive panels). 

For TNG, we do not see prominent differences between the lensing signals around red and blue ICGs. The magenta curves seem to be only slightly above the cyan curves at $10.2<\log_{10}M_\ast/\msun<11.1$.
The trend remains the same if we use true halo central galaxies in TNG300. Besides, we have explicitly checked that TNG100 predicts very similar results. The differences between the magenta and cyan curves for TNG results are mostly not significant. Compared with the errors, TNG red and blue ICGs are hosted by dark matter halos with more similar masses.

Theoretically, it is expected that red galaxies are hosted by more massive halos than blue galaxies with the same stellar mass \citep[e.g.][]{2012ApJ...757....4P,2012MNRAS.424.2574W}. This is because the star formation activities are quenched earlier for red galaxies, but their host dark matter halos keep growing by accretion. Thus at fixed host halo mass, blue galaxies are more massive in stellar mass. In other words, if the stellar mass is the same, red galaxies are hosted by more massive dark matter halos. This is seen in our observed lensing signals, but TNG does not show prominent difference between the lensing signals around red and blue ICGs.

In the following subsections, we move on to compare the satellite and central profiles between real observation and TNG predictions. We will provide more discussions on the difference in host halo mass of red and blue ICGs in Section~\ref{sec:assemble}, by looking at the halo and stellar mass growth histories. 

\begin{figure*}
    \begin{center}
	\includegraphics[width=0.9\textwidth]{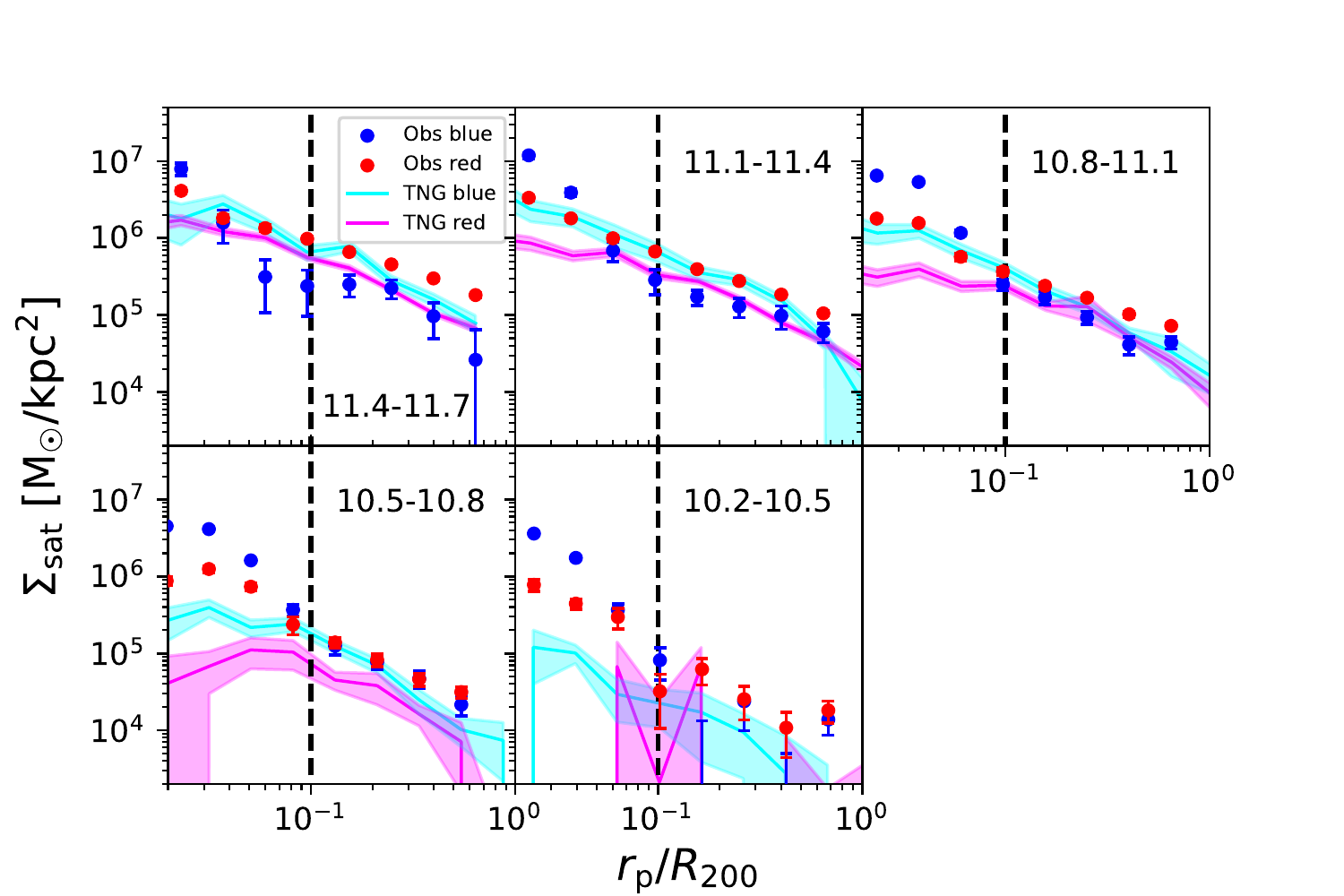}
 \caption{Satellite profiles for observed blue (blue dots) and red (red dots) ICGs, and TNG300 blue (cyan curves) and red (magenta curves) ICGs. Only satellites more massive than $10^9\msun$ are included. Each panel corresponds to a different stellar mass bin, with the text in each panel showing the stellar mass range in unit of $\log_{10}\msun$. Shaded areas indicate the errors for TNG300. All errors are calculated as 1-$\sigma$ scatters of 100 bootstrap subsamples. Note that the observed satellite profiles within 0.1$R_{200}$ are significantly affected by source deblending mistakes with the central dominant galaxy, and hence are not robust. This is marked by the black dashed vertical lines. Beyond 0.1$R_{200}$, there are more satellites around red ICGs in the three most massive panels, but the trend is opposite in TNG300. Stellar mass corrections have been included to account for the missing stellar mass in outskirts of massive galaxies in observation.}
	\label{fig:satprof}
	\end{center}
\end{figure*}

\begin{figure*}
    \begin{center}
	\includegraphics[width=0.9\textwidth]{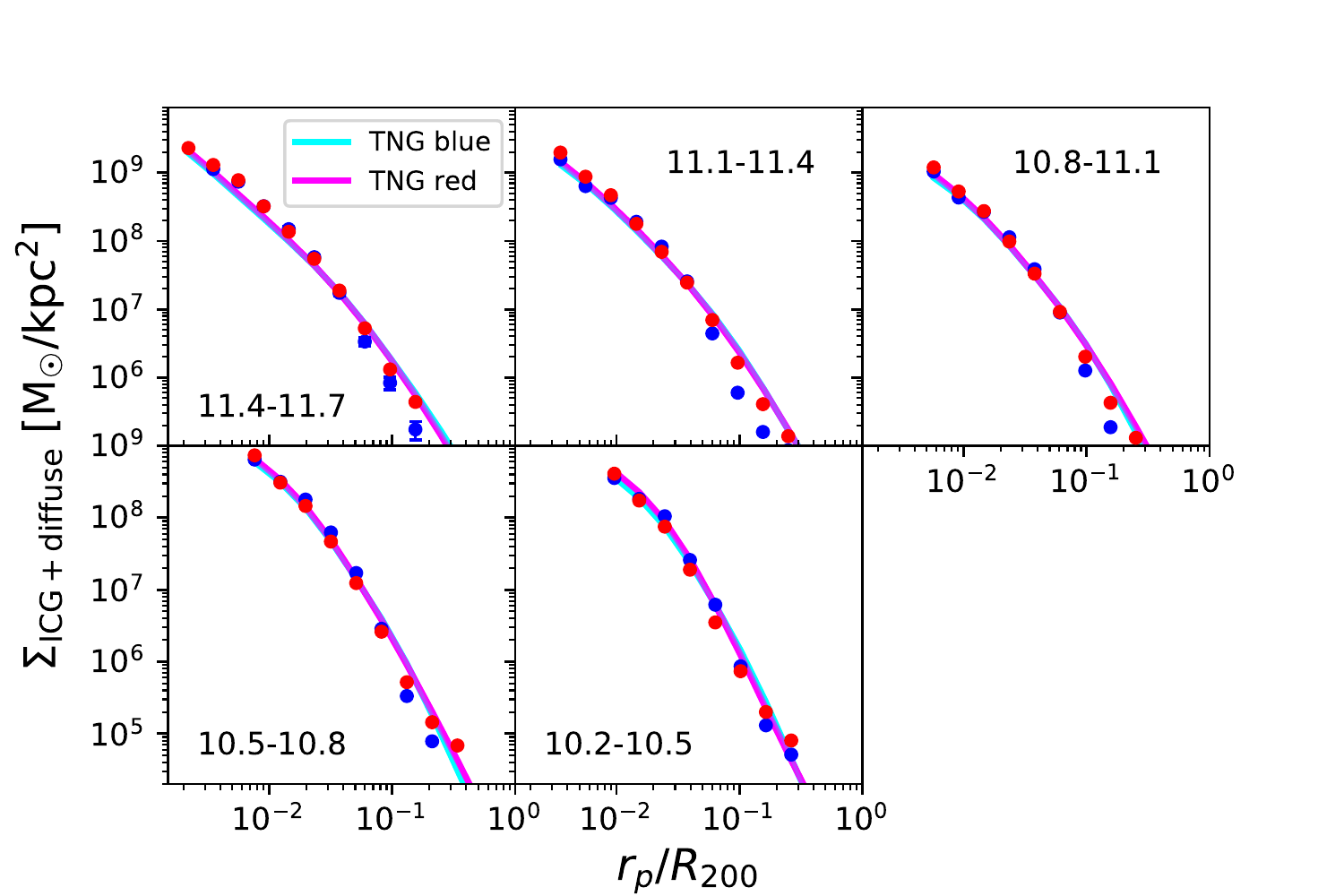}
	\caption{Central profiles for SDSS blue (blue dots) and red (red dots) ICGs, and TNG300 blue (cyan curves) and red (magenta curves) ICGs. Each panel represents the results for a different stellar mass range, and it is indicated by the text in each panel, in unit of $\log_{10}\msun$. All errors are calculated as 1-$\sigma$ scatters of 100 bootstrap subsamples, but they are mostly comparable to the symbol size and line width, hence not prominently shown.}
	\label{fig:cenprof}
	\end{center}
\end{figure*}

\subsection{Satellite profiles}
\label{section:Results_4}

Figure~\ref{fig:satprof} shows the satellite profiles (projected stellar mass density profiles of satellites), for red and blue ICGs separately and for both real observation and TNG300. Although there is a good overall agreement between observation and TNG300, there is a significant disagreement when we look at the profiles around red and blue ICGs separately. In real observation, the satellite profiles are higher in amplitude beyond $0.1R_{200}$ around red ICGs than around the blue ones\footnote{Within $0.1R_{200}$, the observational satellite profiles are strongly affected by deblending mistakes and are thus not robust \citep{2021ApJ...919...25W}. In fact, part of the star-forming regions and spiral arms of blue ICGs are mistakenly deblended as satellites within 0.1$R_{200}$, making the signals artificially higher around blue ICGs at such scales.} for the three most massive panels. This is in good agreement with the previous studies of \cite{2012MNRAS.424.2574W} and \cite{2021ApJ...919...25W}. However, TNG300 shows a different trend, in terms that blue ICGs have satellite profiles higher in amplitudes than the red ones, except for the lowest mass panel.

As having been reported by a few previous studies, the total luminosity or total stellar mass in satellites are good proxies to the host halo mass \citep[e.g.][]{2012MNRAS.424.2574W,2021ApJ...919...25W,2021MNRAS.505.5370T}, and thus we expect the total stellar mass in satellites around red and blue ICGs show similar trends as lensing signals. As we have shown in the previous section, there are indications that red ICGs are hosted by more massive dark matter halos than blue ICGs in real observation, and thus the observed trend in satellite profiles is roughly consistent with what we see in the observed lensing profiles. 

However, there are more stellar mass in satellites around blue ICGs in TNG300, i.e., cyan curves are above magenta ones in Figure~\ref{fig:satprof}, whereas we did not see the cyan curves being prominently above the magenta ones in Figure~\ref{fig:figure_lensing}. Note, to ensure the completeness of satellites in TNG300, the satellite profiles are based on objects which are more massive than $10^9\msun$, which are relatively massive objects, but this does not affect the comparison with observation because we have adopted the same satellite mass threshold for both observation and TNG300. 

We explicitly repeat our calculation for TNG100, and the results are shown in Appendix~\ref{Appendix:Appendix_C}. TNG100 produces results which are noisier due to the smaller simulation box size, but the trend remains very similar, i.e., there are more stellar mass in satellites around blue ICGs. Besides, if using only true halo central galaxies or true satellites bound to central galaxies to calculate their satellite profiles, the trend stays very similar as well. So the inconsistency with observation cannot be trivially explained by the resolution limit or by satellite contamination. 

We have also tried different $g-r$ color divisions when separating ICGs into red and blue populations, such as shifting the division slightly, moving up and down, finding the division according to the dip in the $g-r$ color distribution of each stellar mass bin, and we end up with very similar trends in TNG satellite profiles. 

We postpone more detailed discussions on possible explanations to the discrepancies in Section~\ref{sec:disc}. To gain more evidences, we move on to investigate the central profiles in the next subsection. Moreover, we will investigate in Section~\ref{sec:darkvslum} the fraction of total stellar mass versus total mass, as a function of the projected radius. 

\subsection{Central profiles}
\label{section:Results_5}

Figure~\ref{fig:cenprof} presents the central profiles (projected stellar mass density profiles of ICGs and their extended stellar halos), for both real observation and TNG300. The observed central profiles are directly taken from the measurements of \cite{2021ApJ...919...25W}. TNG300 central profiles exhibit a good overall agreement with real observations. Nevertheless, we observe the TNG300 profiles are more extended, which is particularly relevant for the radial range of $r_\mathrm{p}\approx0.1-0.3R_{200}$ and for blue ICGs more massive than $10^{10.8}\msun$. Similar trend also exists in the $10.5<\log_{10}M_\ast/\msun<10.8$ panel, but not as significant. Note that for the observed central profiles, corrections to the stellar mass (see Section~\ref{section:masscorr}) are not applied. However, this cannot explain the discrepancy. In fact, as we have mentioned, without the correction to the observed stellar mass, we end up selecting more massive galaxies with more extended stellar halos in each bin and in observation. Hence if including corrections to the stellar mass, we would select less massive galaxies in observation, and since less massive galaxies have less extended stellar halos, the discrepancy between real observation and TNG300 predictions would be even larger. We have also explicitly repeated our measurements with TNG100, and have found similar disagreement with real observation.

The more extended central profiles of TNG300 galaxies might be partly related to the fact that the sizes of TNG galaxies are larger than real observed galaxies. It was shown by \cite{2018MNRAS.473.4077P} that the average sizes of TNG galaxies are larger than those of real galaxies. The mass-size relation predicted by TNG cannot even touch the 1-$\sigma$ error boundaries of SDSS galaxies. Part of the large discrepancy could be due to the fact that the outskirts of massive galaxies might fall below the detection limit of the shallower SDSS survey. Moreover, the SDSS mass-size relation measured by \cite{2003MNRAS.343..978S} is based on the circularized radius, which is on average $\sim$1.4 times smaller than the actual major axis length \citep{2017MNRAS.465..722F}. Nevertheless, the fact that hydrodynamical simulations such as TNG and EAGLE tend to predict larger galaxy sizes than observation seems to be a universal problem \citep[e.g.][]{2019MNRAS.486.3702S,2021arXiv211004434Y,2022MNRAS.511.2544D,2022arXiv220307491V}. 

In addition to the connection to the global mass-size relation, our results show that the discrepancy of the TNG predicted central profiles from real observed galaxies is mainly in outskirts, where the stellar material is dominated by accreted stars from satellite galaxies \citep[e.g.][]{2016MNRAS.458.2371R}. This is in good agreement with \cite{2021MNRAS.500..432A}, in which a comparison of the most massive galaxies were performed between TNG100 and the measurements based on the Hyper Suprime-Cam (HSC) imaging data at $z\sim0.4$. It was reported that the outer stellar halos of observed massive galaxies hosted by halos with virial masses of $\sim10^{14}\msun$ are less extended than TNG100 predictions. 

The story is similar for very nearby Milky-Way-mass galaxies. \cite{2020MNRAS.495.4570M} reported a discrepancy between the surface brightness profiles of Milky-Way-like disk galaxies from the Dragonfly Nearby Galaxies Survey (DNGS) and TNG100 counterparts. The TNG100 galaxies were matched in stellar mass to those of DNGS, and surface brightness calculations showed that the DNGS galaxies have "missing" stellar mass in the outskirts. DNGS galaxies have significantly lower surface brightness profiles beyond $20\mathrm{kpc}$. However, as having been pointed out by \cite{2020MNRAS.495.4570M}, their sample size of nearby disk galaxies is relatively small, so it is likely that the results could have been affected by statistical fluctuations. Nevertheless, it seems that at $z=0$, $z\sim0.1$, and $z\sim0.4$, evidences all exist to prove that TNG tends to predict more extended outer stellar halos than real galaxies. Compared with the smaller sample of galaxies in the local Universe, the sample of ICGs is large enough to ensure good statistics, and \cite{2021ApJ...919...25W} have performed intensive tests on background subtraction and companion source masking. In particular, for blue ICGs in the three massive panels, the disagreement between TNG300 prediction and the \cite{2021ApJ...919...25W} measurement is significant compared with the small errors in Figure~\ref{fig:cenprof}.

In addition to the tension we discussed above, we can observe in real data that the outer stellar halos are more extended around red ICGs, but TNG300 does not show such a trend. There is no significant difference between the outer central profiles around red and blue ICGs in TNG. We have checked the same is true in TNG100.

\subsection{Total stellar mass versus total mass}
\label{sec:darkvslum}

\begin{figure*}
    \begin{center}
	\includegraphics[width=0.9\textwidth]{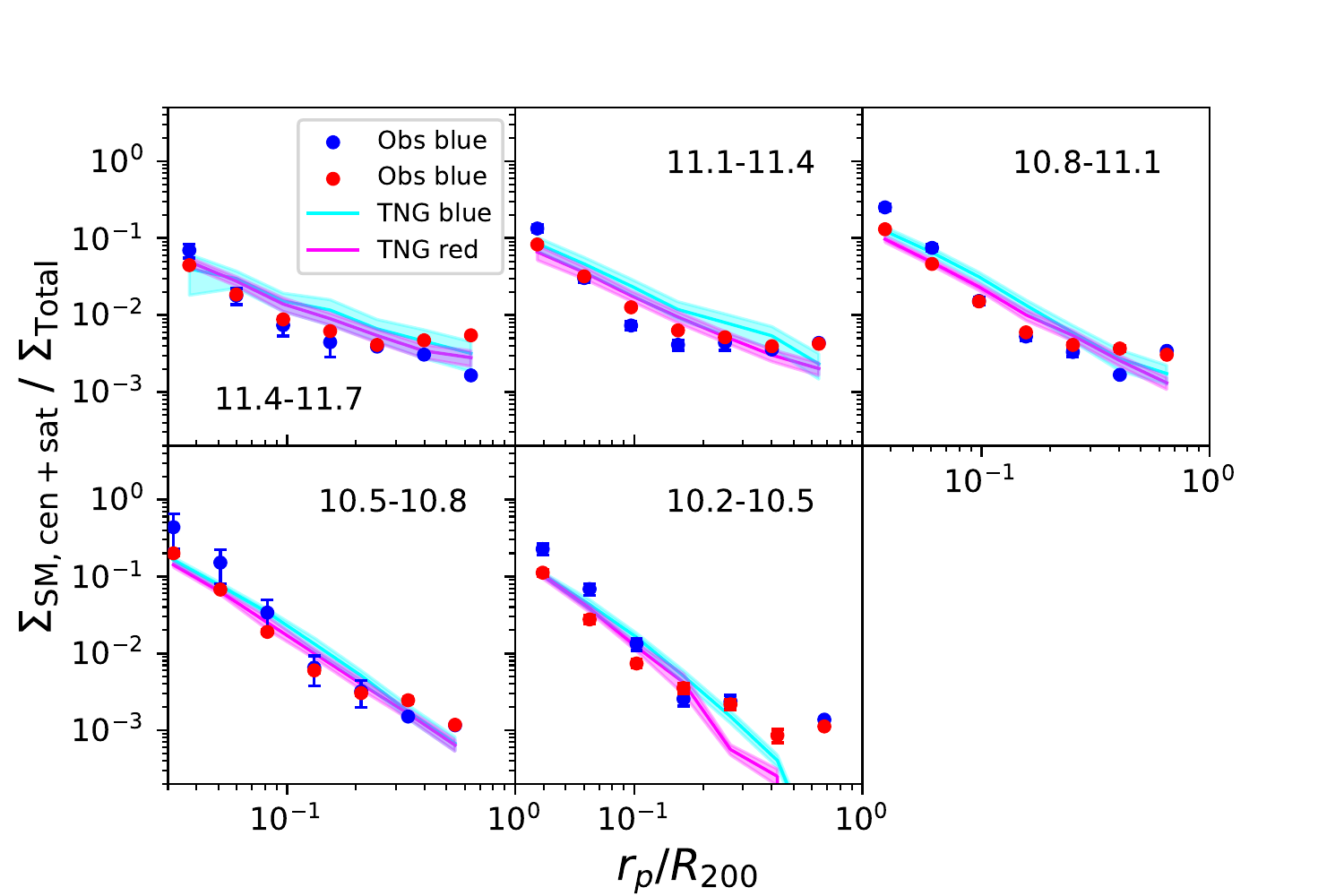}
	\caption{Ratios between the projected density profiles of total stellar mass (central $+$ satellites) versus total mass. Red and blue dots with errors are centered on observed red and blue ICGs, while magenta and cyan curves are around red and blue ICGs in TNG, respectively. Shaded areas indicate the errors for TNG300. The stellar mass range is indicated in each panel, in unit of $\log_{10}\msun$. Errors are propagated from the corresponding central/satellite profiles (Figures~\ref{fig:satprof} and \ref{fig:cenprof}) and lensing signals (Figure~\ref{fig:figure_lensing}). For TNG300 predictions, the projected density profiles of total mass are estimated from the radial distribution of all types of particles. }
	\label{fig:figure_ratio}
	\end{center}
\end{figure*}

\begin{figure*}
    \begin{center}
	\includegraphics[width=0.9\textwidth]{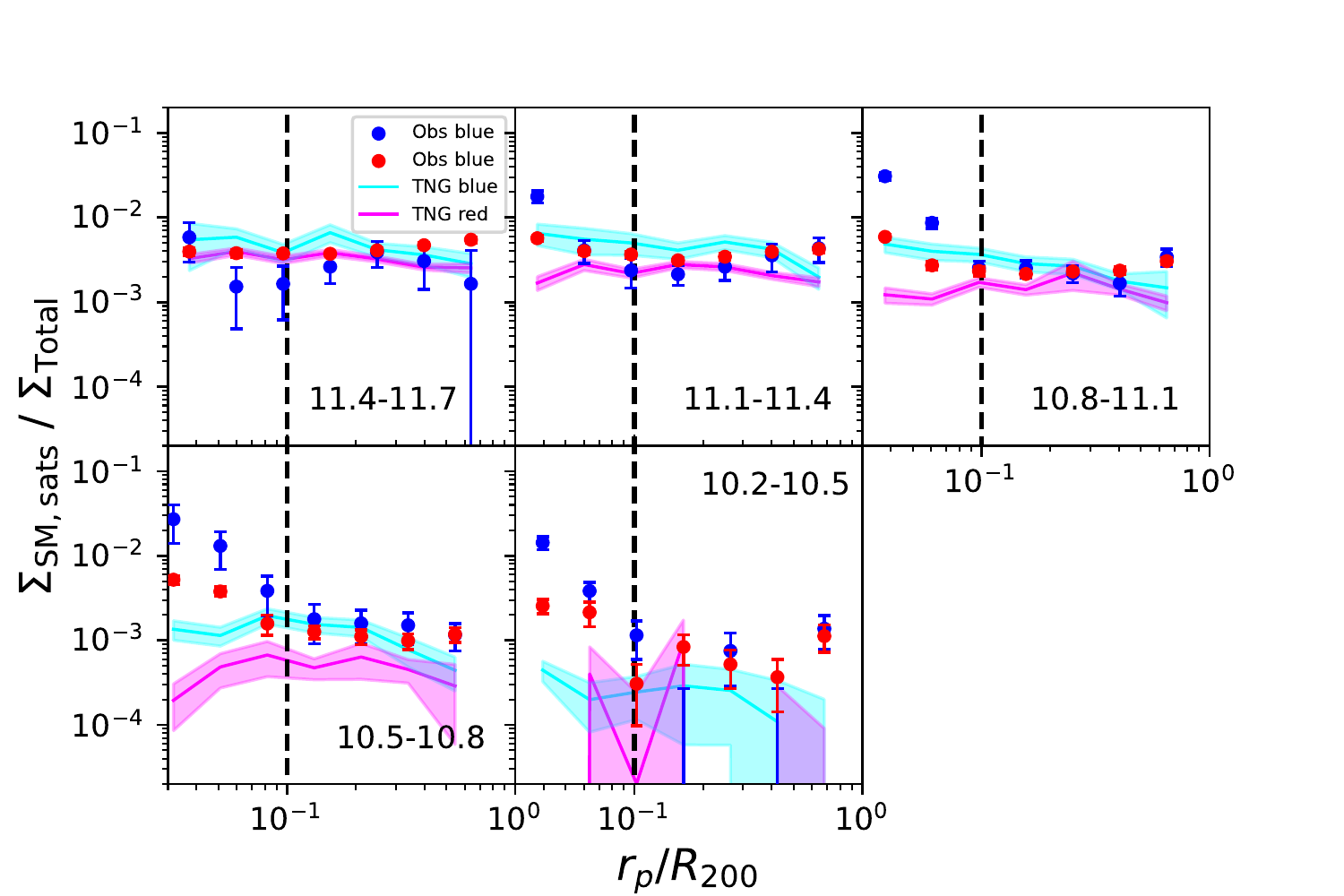}
	\caption{Similar to Figure~\ref{fig:figure_ratio}, but showing the ratios between the projected stellar mass density profiles of satellites versus total mass. Red and blue dots with errors are centered on observed red and blue ICGs, while magenta and cyan curves are around red and blue ICGs in TNG, respectively. Shaded areas indicate the errors for TNG300. Errors are propagated from the corresponding satellite profiles (Figure~\ref{fig:satprof}) and lensing signals (Figure~\ref{fig:figure_lensing}). The black dashed lines mark 0.1$R_{200}$ within which the deblending errors are significant, and therefore the observed satellite profiles are not robust.}
	\label{fig:figure_ratio2}
	\end{center}
\end{figure*}

In this section we investigate how the fraction of total stellar mass versus total mass changes as a function of the projected distance to the galaxy center, for the total stellar mass in central ICGs $+$ their stellar halos and satellites. We also check whether the radial distribution of satellite galaxies trace the underlying dark matter. 

In order to obtain the projected density profiles for the total mass distribution from the observed lensing signals in Figure~\ref{fig:figure_lensing}, we fit the following projected NFW profile \citep{2000ApJ...534...34W} to the observed lensing signals for red and blue ICGs

\begin{equation}
    X_{ij}=\left\lbrace
\begin{aligned}
&\frac{r_\mathrm{s}\delta_\mathrm{c}\rho_\mathrm{c}}{\Sigma_\mathrm{c}}g_<(x)&(x<1)\\
&\frac{r_\mathrm{s}\delta_\mathrm{c}\rho_\mathrm{c}}{\Sigma_\mathrm{c}}\left[\frac{10}{3}+4\mathrm{ln}\left(\frac{1}{2}\right)\right]&(x=1)\\
&\frac{r_\mathrm{s}\delta_\mathrm{c}\rho_\mathrm{c}}{\Sigma_\mathrm{c}}g_>(x)&(x>1)
\end{aligned}
\right.
\end{equation},

where $r_\mathrm{s}=R_{200}/c$ is the scale radius, and $x=r_\mathrm{p}/r_\mathrm{s}$ is the projected radius to galaxy center divided by the scale radius. The functions of $g_<(x)$ and $g_>(x)$ take the following form

\begin{equation}
\begin{aligned}
    g_<(x)=&\frac{8\mathrm{arctanh}\sqrt{\frac{1-x}{1+x}}}{x^2\sqrt{1-x^2}}+\frac{4}{x^2}\mathrm{ln}\left(\frac{x}{2}\right)\\&-\frac{2}{(x^2-1)}+\frac{4\mathrm{arctanh}\sqrt{\frac{1-x}{1+x}}}{(x^2-1)(1-x^2)^{1/2}},
\end{aligned}
\end{equation}
\begin{equation}
\begin{aligned}
    g_>(x)=&\frac{8\mathrm{arctan}\sqrt{\frac{x-1}{1+x}}}{x^2\sqrt{x^2-1}}+\frac{4}{x^2}\mathrm{ln}\left(\frac{x}{2}\right)\\&-\frac{2}{(x^2-1)}+\frac{4\mathrm{arctan}\sqrt{\frac{x-1}{1+x}}}{(x^2-1)^{3/2}}.
\end{aligned}
\end{equation}

For ICGs in TNG300, the projected density profiles for the total mass distribution are directly calculated from the density profiles of all different types of particles, including star and dark matter particles, gas cells, and black holes.

We show in Figure~\ref{fig:figure_ratio} the fractions of total stellar mass out of total mass, for both real observation and TNG300 predictions and as a function of the projected radius scaled by $R_{200}$. We stop at an inner radius of $0.03R_{200}$ of the $x$-axis, because for real observation we do not have weak lensing signals measured on such small scales in Figure~\ref{fig:figure_lensing}, due to the lack of enough lens-source pairs at very small radii. The total stellar mass is estimated from the contribution of both satellite and central profiles, and thus it includes the contribution from surviving satellites, central ICGs and their extended stellar halos formed by disrupted satellites. Note although the satellite profiles are affected by deblending mistakes within $0.1R_{200}$, the total stellar mass is dominated by central ICGs within $0.1R_{200}$, and thus is not affected by deblending mistakes. Besides, for satellites, we only calculate the stellar mass in those which are more massive than $10^9\msun$, so in principle we should take our fractions of total stellar mass as lower limits. However, as long as the power-law index of the satellite stellar mass function does not rise steeper than $-2$ at the low-mass end, we expect the total stellar mass in satellites to be dominated by the most massive/luminous satellites \citep[e.g.][]{2016MNRAS.459.3998L,2021MNRAS.500.3776W}, so we do not expect our fraction to be too different from that of using all satellites. 

In the three most massive panels of Figure~\ref{fig:figure_ratio}, the TNG300 curves are higher than real observations around $r_\mathrm{p}\sim0.15R_{200}$, indicating there are slightly more total stellar mass in TNG300 than real galaxies at $r_\mathrm{p}\sim0.15R_{200}$ around ICGs more massive than $\log_{10}M_\ast/\msun\sim10.8$. This discrepancy is mainly due to the more extended central profiles in TNG compared with real observed galaxies, as the readers can see from Figure~\ref{fig:cenprof} that the TNG central profiles show the most prominent differences from observation at $\sim0.15R_{200}$ and in the three most massive panels.

Besides, we can see in Figure~\ref{fig:figure_ratio} that cyan curves are above magenta ones at almost every radii and in every panels. In outskirts close to the virial radius boundary, the total stellar mass is contributed more by those from surviving satellites, and thus the trend is related to the fact that we see more stellar mass in satellites around blue ICGs in TNG (Figure~\ref{fig:satprof}). In more central regions, this is mainly because there are slightly more total mass around red ICGs than blue ones in TNG (see the TNG300 lensing signals in Figure~\ref{fig:figure_lensing}), which increases the denominator and brings down the fraction in total stellar mass. For the most massive panel, the central profile around blue ICGs are slightly higher, though this is not very prominently seen in Figure~\ref{fig:cenprof}. However, we do not see the same trend in real observation. Blue dots are above red dots in only central regions, and in fact red dots are mostly above blue ones at $r_\mathrm{p}>0.1R_{200}$, except for the lowest mass panel. 

In outskirts of Figure~\ref{fig:figure_ratio}, the results are dominated by surviving satellites, and the fractions already drop below 1\%. It seems observed satellites have slightly more extended radial distributions, in terms that the outer most red and blue dots are above the magenta and cyan curves in the three most massive panels. In the least massive panel, the three red/blue dots at large radius are significantly above the magenta/cyan curves, but this could be due to the reason that the TNG300 predicted satellite profiles are too noisy in this least massive panel (see Figure~\ref{fig:satprof}).



To more clearly investigate and separate the contribution from satellites, we now show in Figure~\ref{fig:figure_ratio2} the ratios between the stellar mass in surviving satellites versus the total mass as a function of projected radius to the centers of ICGs. Despite the discrepancy in the few outer most points between observation and TNG, the radial distribution of satellites mildly trace the underlying dark matter in both real observation and simulation. Note within $0.1R_{200}$ (black dashed vertical lines), the observed satellite profiles are significantly affected by deblending mistakes, hence showing some artificial increases. 

Theoretically, we expect the total mass in accreted subhalos and the total accreted dark matter contributing to the growth of host dark matter halos (including the smoothly accreted part not in any clumpy structures) to be proportional to each other. However, subhalos undergo significant tidal disruptions after infall, so the total mass in surviving subhalos is not really expected to trace well the underlying dark matter. Instead, the amount of stellar mass bound to surviving satellite galaxies in the very center of infalling subhalos are expected to be less significantly stripped, except for those satellites fell in very early and are close to the peri-centers. It is thus expected that the total stellar mass in satellites should approximately trace the spatial distribution of dark matter \citep[e.g.][]{2016MNRAS.457.1208H,2018MNRAS.475.4020W,2022ApJ...933..161M,2022ApJ...929..120D}. This is proved in our measurements of real data, and also approximately reproduced by TNG. 

However, due to the fact that in TNG300 there are more stellar mass in satellites around blue ICGs (Figure~\ref{fig:satprof}), the cyan curves in Figure~\ref{fig:figure_ratio2} are significantly higher than magenta curves. This is not seen in real observation, because red ICGs are hosted by more massive dark matter halos while at the same time have more stellar mass in surviving satellites, and thus the trend is not obviously monotonic between red and blue. 

So far, we have raised a few detailed tensions between TNG300 and observation, when looking at host dark matter halos, population of satellites and outer stellar halos around red and blue ICGs separately. Similar tensions also exist between TNG100 and observation. In the next section, we move on to discuss possible explanations to these tensions. 

\section{Discussions}
\label{sec:disc}

In the previous section, we have shown evidences that in real observation red ICGs are hosted by more massive dark matter halos, have more stellar mass in their satellites and have more extended outer stellar halos than blue ICGs. TNG300 predicts in general a good agreement with real observation, but delicate tensions exist when we look at the red and blue ICGs separately. As we have checked, similar tensions still exist with TNG100. 

First of all, we see that red and blue ICGs in TNG are hosted by more similar dark matter halos. Secondly, blue ICGs have more stellar mass in their satellites than red ICGs in TNG. Moreover, the outer stellar halos are similar between red and blue ICGs in TNG, but overall the outer stellar halos are more extended than real galaxies. This is particularly true for blue ICGs more massive than $10^{10.8}\msun$. Due to the more extended outer stellar halos for TNG galaxies, the total stellar mass versus total mass fractions do not show exactly the same radial distributions as real galaxies.

\subsection{The tension between observed and TNG predicted central/satellite profiles}

Based on the fact that TNG galaxies have more extended outer stellar halos, we suggest that satellite galaxies in TNG300 and TNG100 might be disrupted more efficiently than real observed galaxies, which forms the heavier outer stellar halos. This is because the outer stellar halos are dominated by stripped stars from infalling satellites \citep[e.g.][]{2016MNRAS.458.2371R}.

Moreover, faster satellite disruption in TNG may also weaken the correlation between the total stellar mass in satellites and the host halo mass on a secondary order. Red galaxies are in relatively more over-dense regions than blue ones. If in TNG the satellites are accreted on average earlier around red ICGs than blue ones at fixed stellar mass, the satellites around red ICGs could have been disrupted more due to their earlier infall. Besides, we have seen in Figure~\ref{fig:figure_lensing} that red ICGs are hosted by slightly larger dark matter halos at $10.2<\log_{10}M_\ast/\msun<11.1$ than blue ICGs in TNG, so the tidal stripping can be slightly stronger for satellites around these red ICGs in TNG, due to the more massive host halos. If the tidal disruption in TNG happens more efficiently than real Universe, maybe this would cause less stellar mass remained in bound surviving satellites around red ICGs than their blue counterparts, resulting in an opposite trend compared with real observation.

\begin{figure}
    \begin{center}
	\includegraphics[width=0.49\textwidth]{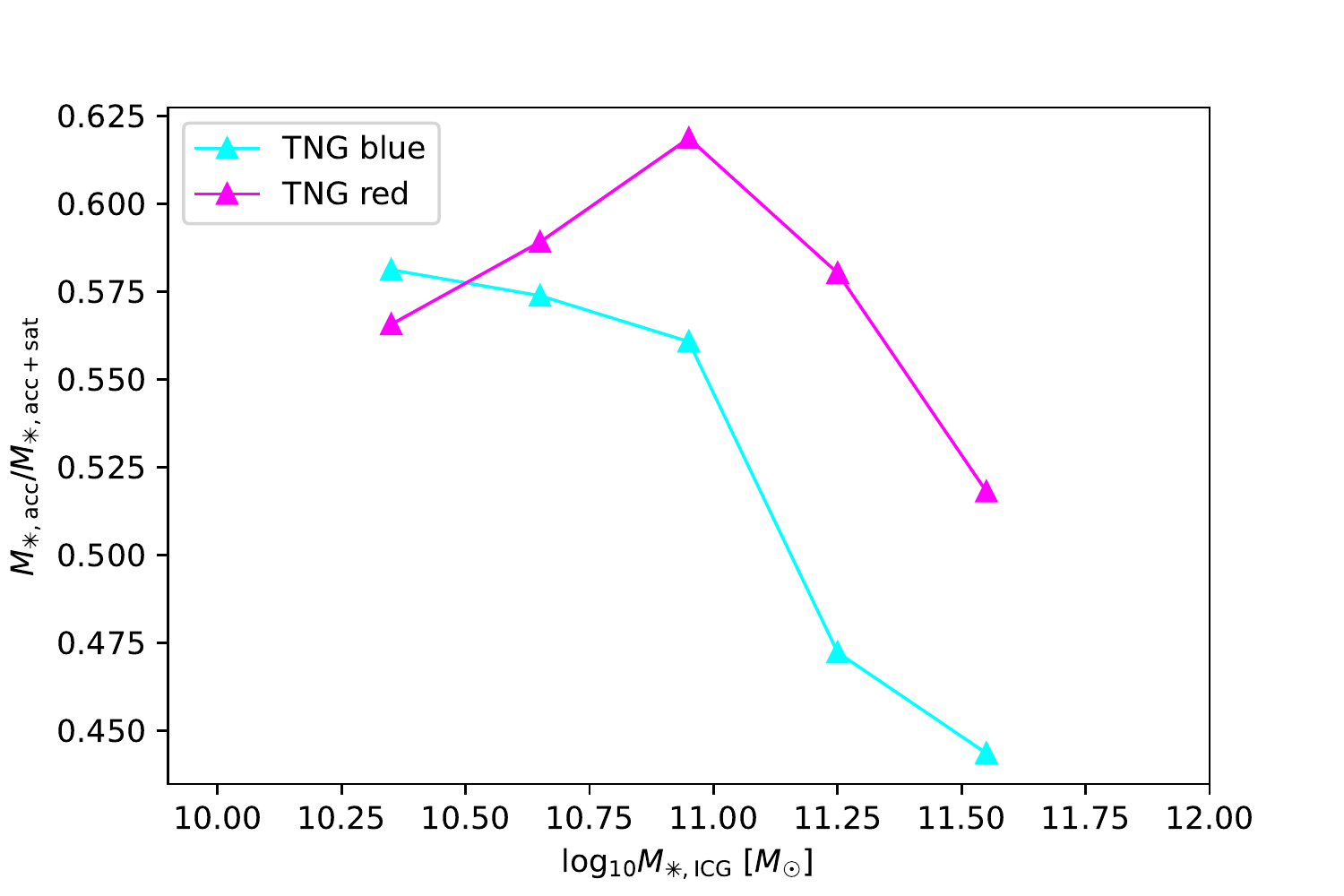}
	\caption{The fraction of ex-situ formed stars from disrupted satellites versus the total accreted stellar mass (disrupted $+$ surviving satellites), for red and blue ICGs in five stellar mass bins.}
	\label{fig:exsitu_frac}
	\end{center}
\end{figure}

We first checked the infall time distribution of satellites around red and blue ICGs, and indeed identified earlier infall of satellites around red ICGs (see Appendix~\ref{Appendix:Appendix_D}). Moreover, the amount of stripped stellar mass from disrupted satellites can be directly known for red and blue ICGs in TNG. Here we use the stellar assembly catalog by \citep{2016MNRAS.458.2371R,Rodriguez-Gomez2017}, which provides the information of ex-situ formed stellar mass for galaxies in TNG. Figure~\ref{fig:exsitu_frac} shows the results for red and blue ICGs in our five stellar mass bins. For the quantity of the $y$-axis, the nominator is the ex-situ formed stellar mass of red and blue ICGs, which includes not only those stripped stars from surviving satellites at $z=0$, but also those entirely disrupted satellites\footnote{In principle, we should only include the stellar mass stripped from $z=0$ surviving satellites, but Figure~\ref{fig:exsitu_frac} here already demonstrates a useful trend.}. The denominator is the ex-situ formed stellar mass plus the total stellar mass in all\footnote{We only count satellites which are more massive than $10^9\msun$ in the satellite profiles, but since massive satellites dominate over stellar mass, if we exclude the stellar mass smaller satellites, the trend of Figure~\ref{fig:exsitu_frac} remains very similar.} $z=0$ surviving satellites within $R_{200}$. It is prominent that the fraction of stripped stars is higher for satellites around red ICGs, except for the least massive bin, so indeed satellites around red ICGs are disrupted faster than those around blue ICGs in TNG for the four massive bins. For the least massive bin the cyan triangle is above the magenta one, while we did not see the cyan curve to be significantly above the magenta one in the corresponding panel of Figure~\ref{fig:satprof} either. We think this is because tidal effects are weaker around less massive ICGs, so satellites around red ICGs are no longer strongly stripped in the least massive bin. Besides, the satellite profile is already quite noisy in the least massive panel of Figure~\ref{fig:satprof}. At least, Figure~\ref{fig:exsitu_frac} and Figure~\ref{fig:satprof} are fully consistent. Thus our explanation that faster satellite disruption in TNG may turn red ICGs to have less stellar mass in their surviving satellites seems plausible.

Note, however, in addition to the fast disruption of satellites around red ICGs, cyan curves in Figure~\ref{fig:satprof} are higher in amplitudes than blue dots in the two most massive bins, indicating satellites around blue ICGs in TNG are NOT stripped enough. However, given the fact that the TNG central profiles are more extended in ourskirts, especially around blue ICG, it seems unlikely that satellites around blue ICGs are not stripped enough, because this should lead to less extended outer stellar halos around blue ICGs. In fact, \cite{2020MNRAS.495.4570M} have shown in their results that artificially delaying the disruption of satellite galaxies and reducing the spatial extent of in-situ stellar populations result in improved matches between the outer surface brightness profile shapes and stellar halo masses. We think the more efficient satellite disruption around red ICGs in TNG is a self-consistent explanation. 

The faster satellite disruption in TNG is at least partially related to the resolution limit. For example, \cite{2021MNRAS.503.4075G} investigated how subhalo mass functions, number density profiles, and substructure mass fractions are affected by mass resolution limits. It was shown that the resolution limit plays a primary role, and lower resolution simulations have more significant disruptions than those in higher resolution simulations, though the conclusions are based on N-body simulations. Moreover, to correct for possible biases introduced by the resolution limit, a rescaling method \citep{2018MNRAS.475..648P} has been adopted to recalibrate the stellar mass in TNG300, which is based on the ratio of stellar mass between two higher resolution TNG100 runs and at fixed halo mass. The stellar mass versus halo mass relations and the stellar mass functions in TNG300 show large differences before and after the recalibration.

Another possible explanation to why satellite profiles are higher in amplitudes around blue ICGs might be related to the galactic conformity phenomenon. In both observation and simulation, people have long been noticed the so-called galactic conformity phenomenon, that the color distribution of satellites correlate with the central galaxies, i.e., satellites around blue centrals are bluer and have stronger star formation rates than satellites around red centrals \citep[e.g.][]{2006MNRAS.366....2W,2006MNRAS.369.1293Y,2012MNRAS.424.2574W,2013MNRAS.430.1447K,2017MNRAS.469.2626H}. This might explain why there are more stellar mass in surviving satellites around blue ICGs, because more stellar mass could have formed in satellites around blue ICGs. Nevertheless, if we want to explain this using the galactic conformity phenomenon, the conformity signal has to be prominently stronger in TNG than real observation. Based on a previous study of \cite{2022arXiv220702218A}, the TNG300 conformity signal is in a general good agreement with SDSS and DESI galaxies. We have repeated the calculation of our conformity signal, based on the sample of ICGs in TNG and observation, and we did not find the conformity signal to be significantly stronger in TNG than that of real galaxies. Thus galactic conformity is unlikely to explain the tension in satellite profiles between TNG and real observation.

\subsection{Indications from the mass growth histories of ICGs in TNG}
\label{sec:assemble}

\begin{figure*}
    \begin{center}
	\includegraphics[width=0.9\textwidth]{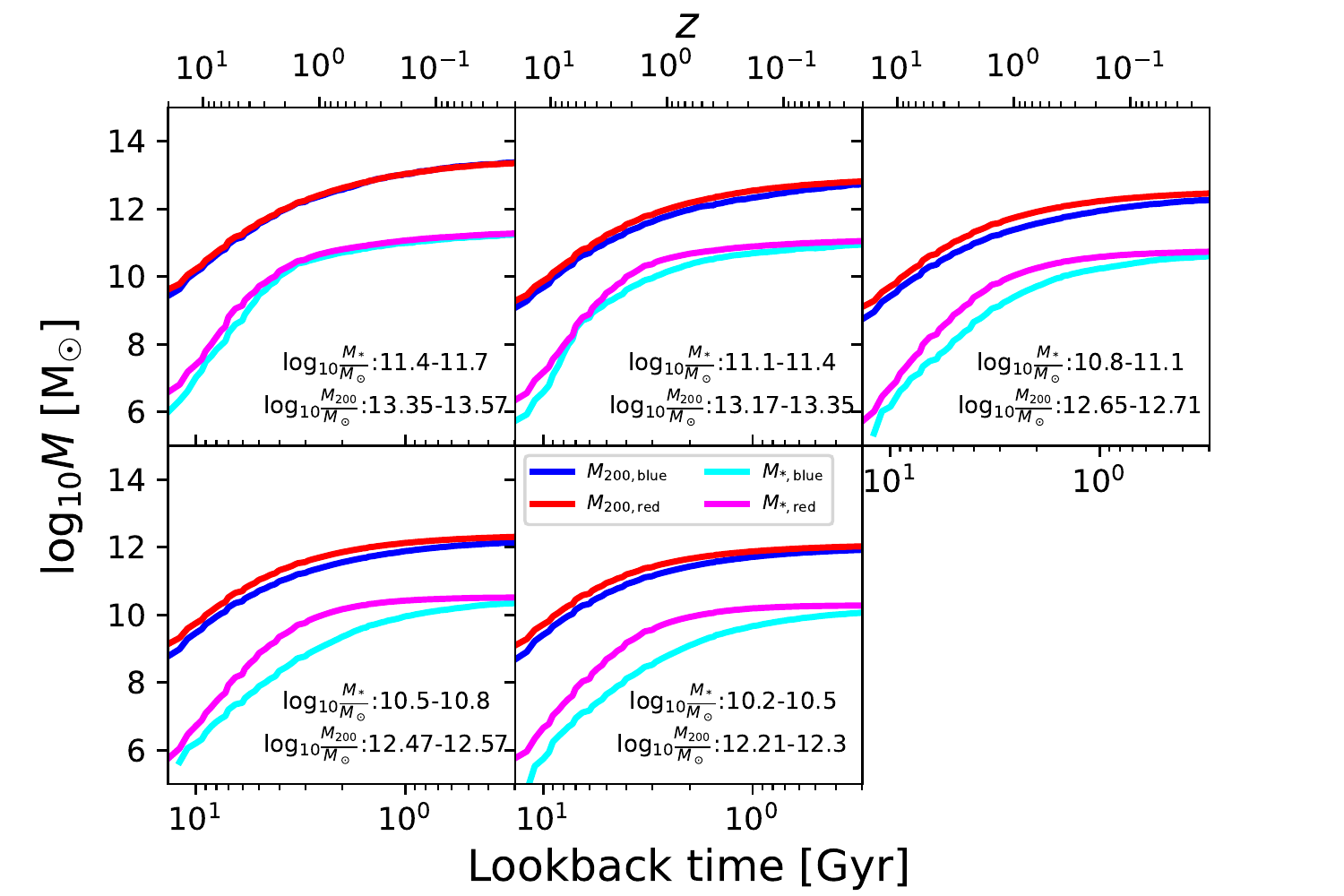}
	\caption{The growth history in stellar mass and $M_{200}$ for TNG300 ICGs. Red and blue curves show the $M_{200}$ growth histories in bins of $\log_{10}M_\ast$, while magenta and cyan curves show the stellar mass growth histories in bins of $\log_{10}M_{200}$, for red and blue ICGs, respectively. Each panel represents the results for a different stellar mass or $M_{200}$ bin, indicated by the text at the bottom of each panel, which is in unit of $\log_{10}\msun$. Note the chosen range of $M_{200}$ bins approximately corresponds to the common range in $M_{200}$ for red and blue ICGs in the corresponding stellar mass bin. }
	\label{fig:figure_merger_history}
	\end{center}
\end{figure*}

As shown in Figure~\ref{fig:figure_lensing}, the lensing signals around red and blue ICGs already do not show exactly the same trend between observation and TNG. As we have mentioned above, the difference between the lensing signals around red and blue ICGs is related to the stellar and halo mass growth histories. Thus in this section we investigate the stellar mass and dark matter growth histories for red and blue ICGs. 

Figure~\ref{fig:figure_merger_history} shows the average growth history of $M_{200}$ for red and blue ICGs in our five stellar mass bins (red and blue curves) and in TNG300. In addition, we also choose five bins in $M_{200}$, which roughly corresponds to the common range in $M_{200}$ for red and blue ICGs in the five stellar mass bins. After selecting red and blue ICGs according to $M_{200}$, the magenta and cyan curves in each panel of Figure~\ref{fig:figure_merger_history} also shows the growth history in stellar mass for these ICGs in $M_{200}$ bins. The ranges in both log stellar mass and in log host halo mass are indicated by the text in each panel. 

At fixed $M_{200}$, red ICGs indeed stop their growth in stellar mass earlier, with the magenta curves starting to become flat\footnote{We define the redshift when the curve starts to become flat, as the time when the change rates in stellar mass start to be smaller than 25\% per Gyr.} at around $z\sim1.6$ for all stellar mass bins, whereas the cyan curves start to be flat at  $z<1.6$, $z<1.07$, $z<0.58$, $z<0.38$ and $z<0.38$, from the most to the least massive bins. The stop in the stellar mass growth for blue ICGs happens later than red ICGs by approximately 1.54, 4.01, 5.51, and 5.51 Gyr on average, from second most to least massive bins. For the most massive bin, the number of blue ICGs is not enough, but the difference is small between magenta and cyan curves in that bin anyway. On the other hand, the growth histories of $M_{200}$ at fixed stellar mass are more similar between red and blue ICGs. The red curves are at most slightly above the blue ones at $z<0.8$ and $10.2<\log_{10}M_\ast/\msun<11.1$, with a bit larger differences at $z>1$. 

In real observation, Figure~\ref{fig:figure_lensing} above has shown that red ICGs are hosted by more massive dark matter halos in the $10.8<\log_{10}M_\ast/\msun<11.1$ panel, and there are similar trends in a few other panels, but with low statistical significances. On the other hand, TNG shows no prominent differences between the lensing signals around red and blue ICGs. 
We propose that if red TNG ICGs can stop their growth in stellar mass earlier, their $z=0$ stellar mass would become smaller, and we would select more massive red ICGs at $z=0$. As a result, the contrast in the host halo mass for red and blue ICGs would become larger at fixed stellar mass. This would bring better agreement with the observed lensing signals around red and blue ICGs, especially in the $10.8<\log_{10}M_\ast/\msun<11.1$ panel.

Based on the mass growth histories for ICGs in TNG300, red ICGs stop the growth in their stellar mass at $z\sim 1.6$ on average, which correspond to 9.78~Gyrs ago. If our argument above is true, nowadays red central galaxies in our real Universe with stellar masses greater than $10^{10.8}\msun$ are expected to stop significant growth in their stellar mass slightly earlier than $\sim$9.78~Gyrs ago, based on the TNG300 model.

\section{Conclusions} 
\label{section:Conclusion}

In this study, we measure the weak lensing signals and the projected stellar mass density profiles of satellite galaxies (satellite profiles) centered on a sample of isolated central galaxies from SDSS DR7 and at $z\sim0.1$. We also take measurements of the projected stellar mass density profiles for these ICGs $+$ their extended stellar halos (central profiles) from a previous study of \cite{2021ApJ...919...25W}. The shear measurements used for weak lensing signal calculations are based on the \textsc{Fourier\_Quad} Method \citep{2019ApJ...875...48Z} and the eighth data release of the Dark Energy Camera Legacy Survey, while the satellite and central profiles are based on photometric sources and galaxy images of the Hyper Suprime-Cam (HSC) survey (pDR3). All the observational measurements are compared in every detail with predictions by IllustrisTNG300, with validations from IllustrisTNG100 to confirm that our comparisons are not affected by the resolution. 

Excellent agreement is found between the lensing signals from real observation and TNG for all ICGs with different stellar mass. In particular, the good agreement at the massive end relies on a correction to the stellar mass of observed massive galaxies, to account for the missing stellar mass in outskirts due to an oversubtraction in the sky background \citep{2013ApJ...773...37H}. Otherwise the observed lensing signals are significantly higher than the TNG predictions.

However, once our sample of ICGs is divided into red and blue populations, we find more detailed tensions between observed and TNG predicted lensing signals. For galaxies with $10.8<\log_{10}M_\ast/\msun<11.1$, red ICGs are hosted by more massive dark matter halos than their blue counterparts in real observation, and similar trends exist in other stellar mass bins, though with low statistical significances. On the other hand, red and blue ICGs in TNG are hosted by more similar dark matter halos. 

Theoretically, the difference in host halo mass for galaxies with the same stellar mass but different color is due to the early quench of star formation in red galaxies, but whose host halos keep growing through accretion, i.e., the stellar mass of red galaxies are smaller at fixed halo mass. We thus propose that if TNG galaxies can stop their growth in stellar mass slightly earlier, the agreement between observed and TNG predicted lensing signals would be better. If our proposal is correct and based on the stellar mass growth histories of TNG300 galaxies, we predict that nowadays red central galaxies in our real Universe with $M_\ast>10^{10.8}\msun$ are expected to stop significant growth in their stellar mass slightly earlier than $\sim$9.78~Gyrs ago.

We find that the satellite profiles around red and blue ICGs do not show a consistent trend with the lensing signals in TNG300. On the contrary, the satellite profiles are higher in amplitudes around blue ICGs than red ones at fixed stellar mass, when real observed satellite profiles are higher in amplitudes around red ICGs. The same is true in TNG100, and is true for true satellites bound to true central galaxies.

The outer stellar halo profiles of galaxies in TNG300 and TNG100, which are dominated by stripped stars from satellites, are more extended than real observation, especially for blue ICGs at $\log_{10}M_\ast/\msun>10.8$. This is in good agreement with previous studies based on more massive elliptical galaxies at $z=0.4$ \citep{2020MNRAS.495.4570M} and based on a smaller sample of Milky-Way-like galaxies in the local volume \citep{2021MNRAS.500..432A}. Moreover, while we see that red ICGs have more extended outer stellar halos than blue ones in observation, we do not see the same trend in TNG, with red and blue ICGs showing similar shapes in their outer stellar halos.

The more extended outer stellar halo profiles of galaxies in TNG300 and TNG100 than those of real galaxies can be explained by a more efficient satellite disruption compared with that in the real Universe. In addition, we identified higher fractions of disrupted stellar mass around red ICGs than around their blue counterparts. If satellite disruptions in TNG happen more efficiently than in real observation, satellites around red ICGs would have been disrupted more than those around real galaxies, which can explain why there are less stellar mass in surviving satellites around red ICGs, in contrary to real observation. 

The more efficient satellite disruption in TNG and the disagreement with real observation are at least partially due to the resolution limit, in terms that lower resolution simulations have more significant satellite disruptions than those in higher resolution simulations \citep[e.g.][]{2018MNRAS.475..648P,2021MNRAS.503.4075G}.

In both real observation and TNG, satellites approximately trace the underlying radial distribution of dark matter beyond $0.1R_{200}$. However, the fractions of total stellar mass versus total mass do not show exactly the same radial distribution as real galaxies. The fraction of total stellar mass is slightly higher around blue ICGs than red ones in TNG at every radius, but we do not see the same trend in real observation.

\section{Acknowledgments}
This work is supported by NSFC (12022307, 12273021, 11621303, 11890691, 12073017), National Key Basic Research and Development Program of China (No. 2018YFA0404504), 111 project (No. B20019), Shanghai Natural Science Foundation (No. 19ZR1466800) and the science research grants from the China Manned Space Project (No. CMS-CSST-2021-A02, No. CMS-CSST-2021-B03, No. CMS-CSST-2021-A01). We gratefully acknowledge the support of the Key Laboratory for Particle Physics, Astrophysics and Cosmology, Ministry of Education. We thank the sponsorship from Yangyang Development Fund.
SS acknowledges support from NSFC grants (Nos.11988101, 12273053), CAS Project for Young Scientists in Basic Research 
Grant (No. YSBR-062), the K. C. Wong Foundation, and the science research grants from the China Manned Space Project 
(No. CMS-CSST-2021-B03).
QG is supported by NSFC (12033008, 11622325), and the science research grants from the China Manned Space Project (No. CMS-CSST-2021-A03 and CMS-CSST-2021-A07).
CNH acknowledges support from the NSFC grant (No. 11733002).

The computations in this paper were run on the $\pi$ 2.0 cluster supported by the Center of High Performance Computing at Shanghai Jiaotong University, and the Gravity supercomputer of the Astronomy Department, Shanghai Jiaotong University.

\bibliography{master}{}
\bibliographystyle{aasjournal}

\clearpage

\appendix

\section{Consistency of the lensing results}
\label{Appendix:Appendix_A}

\begin{figure*}
    \begin{center}
	\includegraphics[width=0.9\textwidth]{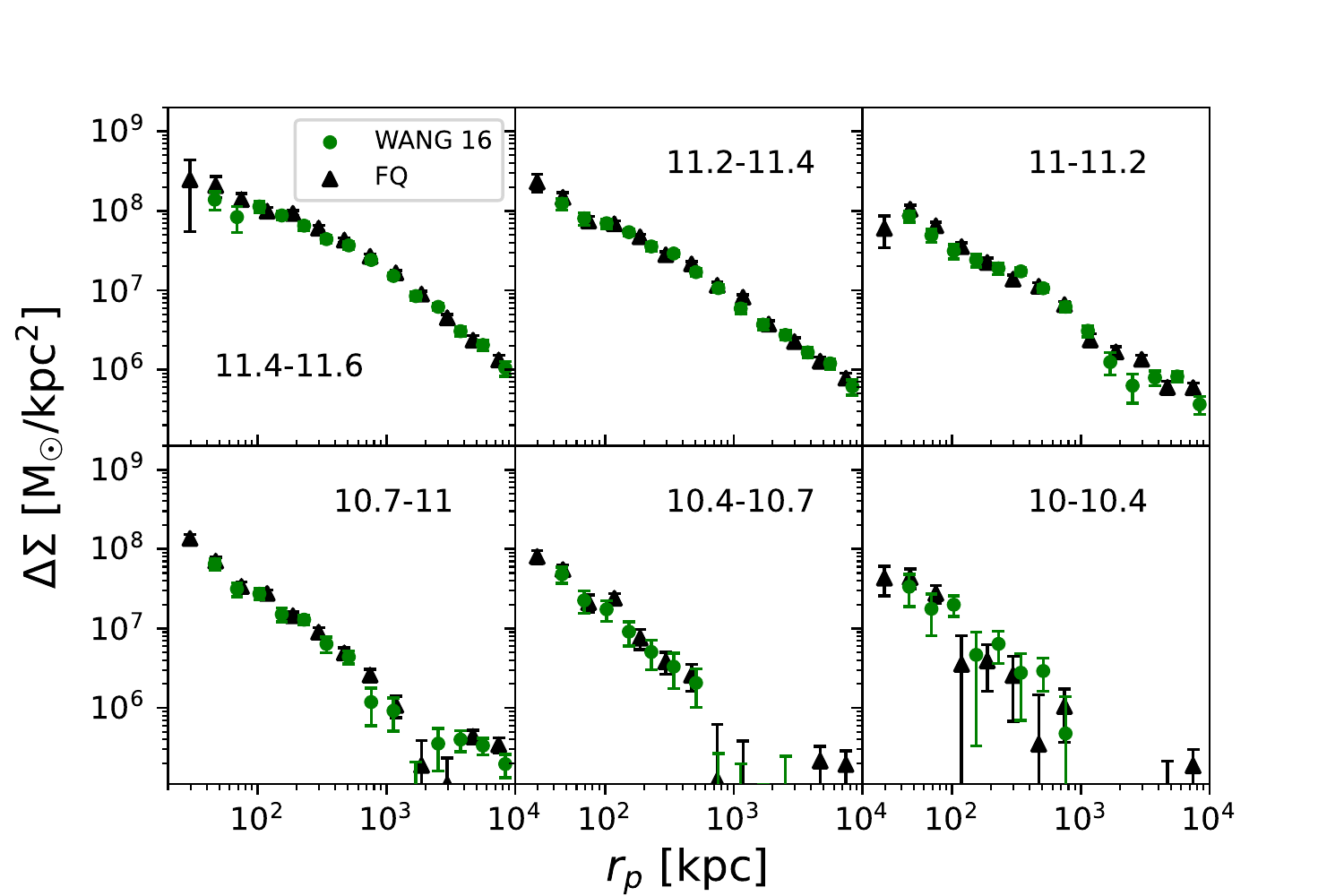}
	\caption{Lensing profiles for locally brightest galaxies (LBGs) selected from SDSS DR7, grouped in different stellar mass bins. Black points show the results obtained by using the \textsc{Fourier\_Quad} pipeline (FQ), using DECaLS data, and green points represent the results from \cite{2016MNRAS.457.3200M} and \citet{2016MNRAS.456.2301W} (Wang 16). The ranges in stellar mass are shown by the text in each panel, in unit of $\log_{10}M_{\odot}$. Note since \cite{2016MNRAS.457.3200M} and \citet{2016MNRAS.456.2301W} did not include corrections to the stellar mass of massive galaxies (see Appendix~\ref{Appendix:Appendix_B} below), no stellar mass corrections have been made to LBGs in this plot, to ensure fair comparisons. }
	\label{fig:lbg}
	\end{center}
\end{figure*}

Figure~\ref{fig:lbg} shows the comparison between the lensing signals calculated using \textsc{Fourier\_Quad} and the results from \cite{2016MNRAS.457.3200M} or \cite{2016MNRAS.456.2301W}. Here we use exactly the same sample of locally brightest galaxies (LBGs) from SDSS DR7 as \cite{2016MNRAS.456.2301W}, and the only difference is the shear catalog. We used photometric sources from DECaLS, whereas \cite{2016MNRAS.456.2301W} used photometric sources from SDSS. Our \textsc{Fourier\_Quad} method is also different from that of the previous studies. Encouragingly, we find a very good agreement between our results and \cite{2016MNRAS.456.2301W}, which allows us to confidently use the \textsc{Fourier\_Quad} method to calculate the lensing signals in this paper. Notably, \cite{2016MNRAS.456.2301W} or \cite{2016MNRAS.457.3200M} did not include any correction to the stellar mass of LBGs, and thus to ensure fair comparisons with previous studies, stellar mass corrections are not included in our results of Figure~\ref{fig:lbg}.

\section{Stellar Mass Correction} 
\label{Appendix:Appendix_B}

\begin{figure*}
    \begin{center}
	\includegraphics[width=0.9\textwidth]{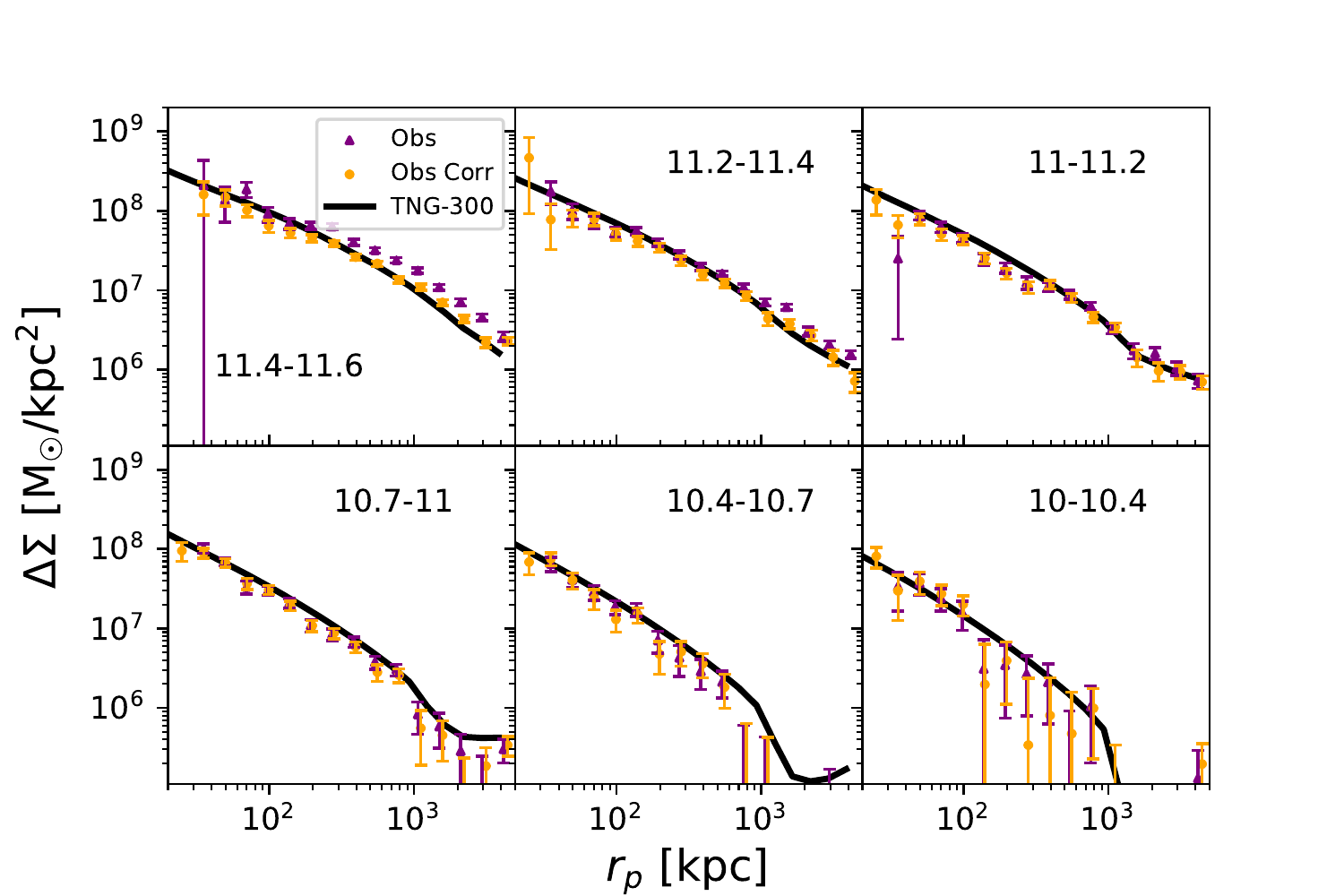}
	\caption{Lensing signals around locally brightest galaxies (LBGs) in six different stellar mass bins. The ranges in stellar mass are shown by the text in each panel, in unit of $\log_{10}M_{\odot}$. Purple points represent the observational results without stellar mass correction. Orange points represent the observational signals after the correction in stellar mass has been applied. The black lines show the TNG300 predictions. Without the stellar mass corrections, the observed lensing signals are significantly overestimated in the two most massive panels. }
	\label{fig:lgbs_after_corr}
	\end{center}
\end{figure*}

\cite{2013ApJ...773...37H} performed a photometric analysis of early-type galaxies (ETGs) from SDSS DR7, with $r$-band magnitude brighter than -22.5, stellar mass greater than $10^{11}\msun$ and redshifts between 0.05 and 0.15. They argue that SDSS underestimated the luminosity due to two possible reasons. Firstly, the PHOTO algorithm of SDSS overestimates the sky background subtraction, which leads to an underestimate of the luminosity. This overestimate of the sky background is mainly due to wrong masking of the galaxy neighbors and extended stellar halos. \cite{2011ApJS..193...29A} raised this problem for SDSS DR7, and introduced an improved sky subtraction algorithm for SDSS DR8, leading to a smaller but still significant underestimate in magnitude. Secondly, the surface brightness distribution of galaxies does not fit well to a Sersic profile (\cite{1963BAAA....6...41S}), and this can also cause underestimates in galaxy luminosity or stellar mass.

\cite{2013ApJ...773...37H} achieved a more accurate photometry for SDSS ETGs by improving the sky background subtraction. In their method, they first used Sextractor \cite{1996A&AS..117..393B} for object masking. They found that, for bright galaxies of $M_\mathrm{r}<-22.5$, a significant number of companion objects were not properly masked, mainly due to the galaxies being present in crowded fields, or having an extended stellar halo, or being near foreground bright stars. They then correct the masks by hand, one by one. This correction leads to a much more accurate photometry, and they measured an underestimate in magnitude by $\sim$0.2-0.3 mag for SDSS DR7.

Following this line of thought, we perform a stellar mass correction (see Section ~\ref{section:masscorr} for details). We then re-calculate the lensing signals with the new stellar mass binning. Figure~\ref{fig:lgbs_after_corr} shows the observational results before and after the correction in stellar mass, together with the simulated results from TNG300. Here we use the same sample of locally brightest galaxies (LBGs) from SDSS DR7 as \cite{2016MNRAS.456.2301W}. We find that the stellar mass correction solves the discrepancy found in the two most massive bins. Without the correction in stellar mass, the observed lensing signals are significantly higher than TNG in the two most massive bins and in outskirts. 

\section{Satellite profiles in TNG100}
\label{Appendix:Appendix_C}

\begin{figure*}
    \begin{center}
	\includegraphics[width=0.9\textwidth]{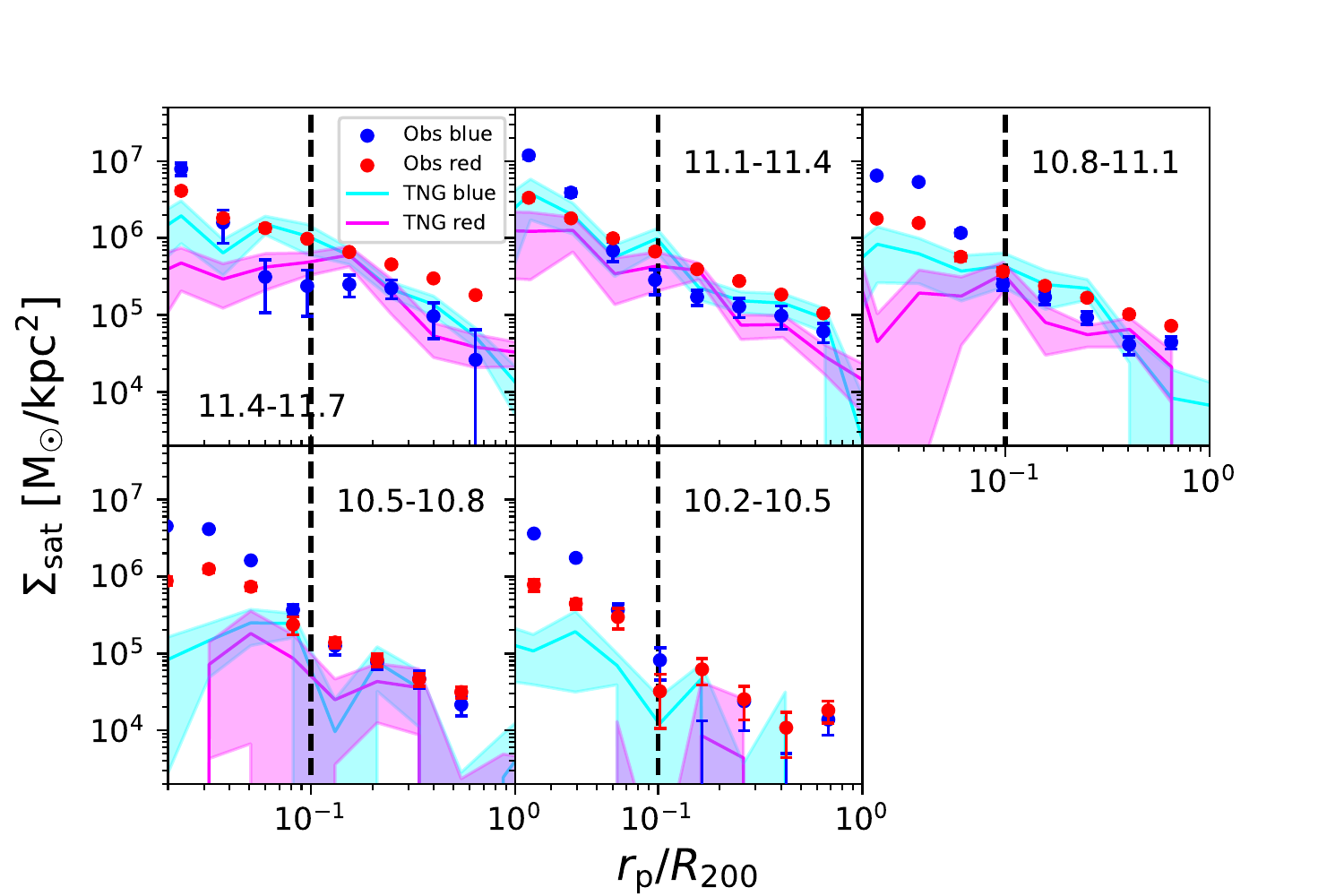}
	\caption{Satellite profiles around observed blue (blue points) and red (red points) ICGs, and TNG100 blue (cyan shaded curves) and red (magenta shaded curves) ICGs, divided in different stellar mass bins, at redshift $z=0$. Satellites more massive than $10^8\msun$ are included to calculate the profiles. The ranges in stellar mass are shown by the text in each panel, in unit of $\log_{10}M_{\odot}$. Errorbars are the 1-$\sigma$ scatters of 100 bootstrap subsamples. }
	\label{fig:sats_TNG100}
	\end{center}
\end{figure*}

Figure~\ref{fig:sats_TNG100} shows the satellite profiles for TNG100 ICGs, compared to real observations. The resolution limit of TNG100 is about one order of magnitude better than TNG300, so we include satellites which are more massive than $10^8\msun$ in Figure~\ref{fig:sats_TNG100}. Similar to the TNG300 results, red and blue ICGs in TNG100 show different trends from those of real observations, though noisier. This discards the resolution limit as a possible explanation to these discrepancies. 

\section{Infall time distribution of satellite galaxies}
\label{Appendix:Appendix_D}

\begin{figure*}
    \begin{center}
	\includegraphics[width=0.9\textwidth]{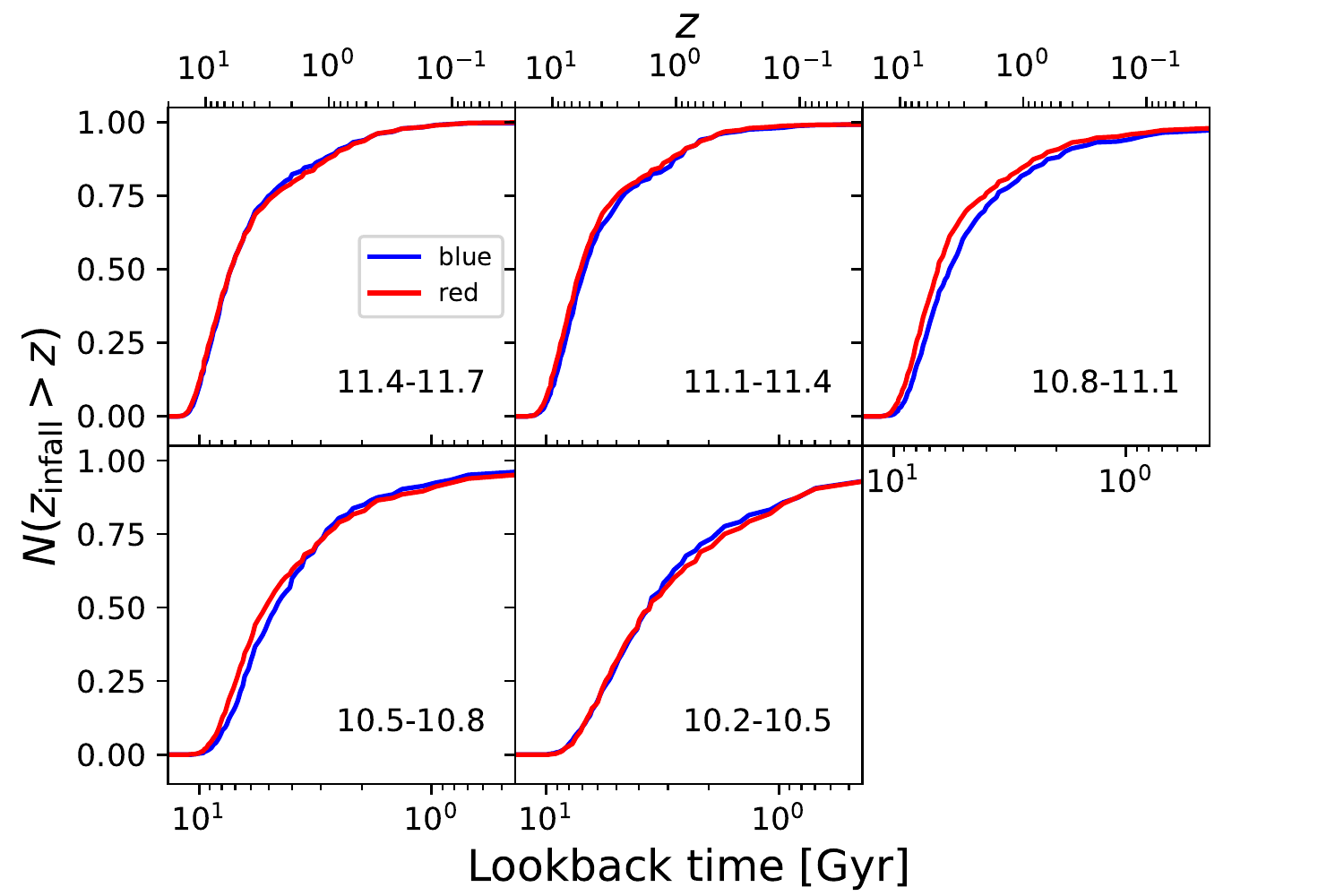}
	\caption{Cumulative distribution of the infall time for satellites around red (red curves) and blue (blue curves) ICGs in TNG300. Each panel represents the results for a different stellar mass bin, indicated by the text at the bottom of each panel, in unit of $\log_{10}\msun$.}
	\label{fig:figure_infall_time}
	\end{center}
\end{figure*}

Figure~\ref{fig:figure_infall_time} shows the cumulative infall time distribution\footnote{We defined the infall time of a $z=0$ satellite as the highest redshift when it was still a satellite galaxy.} of $z=0$ surviving satellites in TNG. We identify some differences between the infall time distribution of satellites around red and blue ICGs. The red curves have slightly earlier infall time distributions than the blue curves in the three middle panels. This indicates that for ICGs with $10.5<\log_{10}M_{\ast}/\msun<11.4$, earlier infall of satellites around red ICGs can partially explain the stronger tidal stripping. However, we also note that the infall time distribution of satellites around red and blue ICGs looks similar in the most massive panel of Figure~\ref{fig:figure_infall_time}. And in the second most massive panel, the difference between red and blue curves is minor. The stronger stripping of satellites around red galaxies in TNG might depend on other factors in addition to the infall time distribution, such as the difference in host halo masses between red and blue ICGs at higher redshifts, the orbits of infalling satellites and how long can satellites maintain their star formation after infall and hence stay more self bound etc.





\end{document}